\documentclass[trackchanges,twocolumn]{aastex701}
\usepackage{booktabs}
\usepackage{makecell}
\usepackage{tikz}
\usepackage{eso-pic}
\usepackage{xcolor}
\usepackage{graphicx}
\usepackage{rotating}
\usepackage{amsmath}
\usetikzlibrary{calc}

\newcommand{\tcell}[2]{%
  \parbox[c]{#1}{\centering #2}%
}

\newcommand{\tlcell}[2]{%
  \parbox[c]{#1}{\raggedright #2}%
}

\begin{document}

\title{The Dancing Bonfire: An Unusual Persistent X-ray Source in M81 Exhibiting Recurrent Luminous Flares}

\author[orcid=0009-0007-2241-7252,sname='Zhang']{Jiachang Zhang}
\affiliation{School of Astronomy and Space Science, Nanjing University, Nanjing 210046, China}
\affiliation{Key Laboratory of Modern Astronomy and Astrophysics (Nanjing University), Ministry of Education, Nanjing 210046, China}
\email[show]{zhangjiachang@smail.nju.edu.cn}  

\author[orcid=0000-0003-0355-6437, sname='Li']{Zhiyuan Li} 
\affiliation{School of Astronomy and Space Science, Nanjing University, Nanjing 210046, China}
\affiliation{Key Laboratory of Modern Astronomy and Astrophysics (Nanjing University), Ministry of Education, Nanjing 210046, China}
\affiliation{Institute of Science and Technology for Deep Space Exploration, Nanjing University, Suzhou 215163, China}
\email[show]{lizy@nju.edu.cn}
\author[orcid=0000-0002-0584-8145, sname='Li']{Xiangdong Li}
\affiliation{School of Astronomy and Space Science, Nanjing University, Nanjing 210046, China}
\affiliation{Key Laboratory of Modern Astronomy and Astrophysics (Nanjing University), Ministry of Education, Nanjing 210046, China}
\email{lixd@nju.edu.cn}
\author[orcid=0000-0002-5082-5049, sname='Bao']{Tong Bao}
\affiliation{INAF - Osservatorio Astronomico di Brera, Via E. Bianchi 46, I-23807 Merate, Italy}
\affiliation{Institut für Physik und Astronomie, Universität Potsdam, Karl-Liebknecht-Str. 24/25, 14476 Potsdam, Germany}
\email{xzbaotong@gmail.com}
\author[orcid=0000-0002-7518-337X, sname='Hua']{Ziqian Hua}
\affiliation{School of Astronomy and Space Science, Nanjing University, Nanjing 210046, China}
\affiliation{Key Laboratory of Modern Astronomy and Astrophysics (Nanjing University), Ministry of Education, Nanjing 210046, China}
\email{602023260006@smail.nju.edu.cn}
\author[orcid=0000-0001-7595-1458, sname='Weng']{Shanshan Weng}
\affiliation{Department of Physics and Institute of Theoretical Physics, Nanjing Normal University, Nanjing 210023, China}
\email{wengss@njnu.edu.cn}

\begin{abstract} 
Short-timescale X-ray variability provides a direct probe of unstable accretion onto compact objects, yet extragalactic X-ray binaries (XRBs) with recurrent flaring remain largely unexplored. 
Here we present a detailed X-ray study of M81F4 (2CXO~J095524.2+690957), a persistently luminous X-ray source towards M81 that exhibits recurrent short X-ray flares. Based on multi-mission X-ray observations spanning two decades, it is found that M81F4 remains persistently bright at $L_{0.5-8.0\rm~keV}\sim10^{38}-10^{39}\rm~erg~s^{-1}$, while producing flares lasting a few hundred seconds and reaching peak luminosities of a few $10^{39}\rm~erg~s^{-1}$, with an observed mean (minimum) recurrence time of $\sim0.5$ day ($\sim$6 ks). The stacked flare profile shows a rapid rise followed by a slower decay, but no significant hardness evolution. Spectral analysis reveals that both the flare and non-flare states are dominated by a hard power-law continuum, with marginal evidence of highly ionized Fe lines in the non-flare state spectrum. These properties disfavor classical Type I and Type II bursts and have no close analog among known XRBs. 
We favor a magnetized neutron star accreting from an intermediate mass donor at a persistently high rate with the flares arising from instabilities at the disk-magnetosphere boundary. If confirmed, M81F4 might represent a hidden population of XRBs in which sustained, magnetically-regulated accretion can lead to brief episodes of super-Eddington emission. 
\end{abstract}


\section{Introduction}

X-ray binaries (XRBs), in which a neutron star (NS) or black hole (BH) accretes from a stellar companion, provide important laboratories to study the profound physics of compact object accretion and binary evolution. 
Short-timescale X-ray variability has long been recognized among Galactic XRBs, which arise from distinct physical mechanisms. Type I X-ray bursts (or thermonuclear bursts)  from accreting NSs are characterized by a rapid rise followed by a cooling decay \citep{2020ApJS..249...32G}, whereas Type II X-ray bursts, in particular those found in two XRBs, the Rapid Burster and GRO~J1744$-$28, are generally associated with instabilities in the accretion disk and disk-magnetosphere interaction \citep{2015MNRAS.449..268B,2018MNRAS.481.2273C}. 

More generally, theoretical models of accretion onto magnetized NSs predict cyclic or episodic accretion instabilities when the inner boundary of the accretion disk approaches the corotation radius \citep{2010MNRAS.406.1208D,2012MNRAS.420..416D}. Possible manifestations of such instabilities have been identified in a small number of Galactic systems, including the Type II bursts and accreting millisecond pulsars, but these occur in transient or relatively low-luminosity accretion regimes. At higher accretion rates, rare systems such as LMC X-4 show recurrent minute-scale super-Eddington flares \citep{2003ApJ...586.1280M}, possibly associated with inhomogeneous accretion onto the NS's magnetic polar caps from a high-mass donor. 
Such systems offer an invaluable opportunity to probe disk-magnetosphere coupling and accretion instabilities under extreme accretion conditions, but they are essentially absent in our Galaxy.
Extragalactic X-ray surveys are particularly valuable in this context, enabling a large population of X-ray binaries with ascertained distances and in diverse stellar environments to be studied systematically, while opening a wider discovery space for rare and exotic accretion-powered systems.

In our recent work based on archival {\it Chandra} observations \citep{2026MNRAS.546f2237Z}, we conducted a systematic search for short-timescale variability among the X-ray sources in three nearby galaxies, M31, Cen A and M81.  
Most remarkable among the sources searched is M81F4 (following the naming convention in \citealp{2026MNRAS.546f2237Z}), which exhibits recurrent flares with durations of a few hundred seconds and peak X-ray luminosities of a few $10^{39}\rm~erg~s^{-1}$. 
Here we present an in-depth X-ray study of M81F4 to address key unresolved questions about this unusual source: what are the basic properties of the flares, what sustain their recurrence, and  what is the physical nature of M81F4. 
To this end, we utilize multi-mission X-ray observations, including {\it Chandra}, {\it XMM-Newton}, {\it Swift}/XRT, and {\it NuSTAR} observations, together with information in the optical. 
The collective properties of M81F4, in particular its persistent high X-ray luminosity and recurrent minute-scale flares, distinguish it from known XRBs. 
We suggest that M81F4 hosts a magnetized NS, accreting from an intermediate-mass donor, with flares arising from instabilities at the disk-magnetosphere boundary.

\section{X-ray Observations and Data Preparation}
\label{sec:data}
M81F4 is identified as the {\it Chandra} source 2CXO J095524.2+690957, which is located at $\sim 6 ^{\prime}$ ($\sim$6.3 kpc) north of the M81 nucleus (see Figure 1 in \citealp{2026MNRAS.546f2237Z}). 
The source has been extensively observed with {\it Chandra}, {\it XMM-Newton}, {\it Swift}/XRT, and {\it NuSTAR}. 
Table~\ref{tab:obs} summarizes the {\it XMM-Newton}, {\it NuSTAR} and {\it Chandra} grating observations used in this work. 
The {\it XMM-Newton}/European Photon Imaging Camera (EPIC) observations were used to search for additional flares and to construct the long-term X-ray light curve. The {\it NuSTAR} observations were used to examine the high-energy spectral behavior of the source. The zeroth-order image data from eight \textit{Chandra}/ACIS grating observations were used to supplement the long-term light curve. The {\it Chandra} non-grating observations used in this work were presented in \cite{2026MNRAS.546f2237Z}. The {\it Chandra} observations analyzed in this work are included in the Chandra Data Collection \dataset[doi:10.25574/cdc.712]{https://doi.org/10.25574/cdc.712}. The {\it Swift} X-ray telescope (XRT) has extensively monitored M81 since its launch, resulting in a total of 486 observations that cover M81F4, spanning 2006-2026. For brevity, information of the individual XRT observations are not listed here, but can be looked up from were obtained from HEASARC.

\begin{table*}[htbp]
    \centering
    \vspace{0.4cm}

    \begin{tabular}{ccccc}
    \hline
    ObsID & Start Time (UTC) & R.A. (J2000) & Dec. (J2000) & Exp. Time (ks) \\
    \hline
    {\it XMM-Newton} &&&&\\
    0111800101 & 2001-04-22 10:04:34 & 09:55:33.00 & +69:03:55.0 & 132.7 \\
    0657802001 & 2011-03-24 15:40:14 & 09:56:43.20 & +69:03:50.4 & 27.5 \\
    0657801601 & 2011-04-17 15:08:44 & 09:56:43.20 & +69:03:50.4 & 21.1 \\
    0657801801 & 2011-09-26 04:17:58 & 09:56:43.20 & +69:03:50.4 & 25.4 \\
    0657802201 & 2011-11-23 00:41:51 & 09:56:43.20 & +69:03:50.4 & 23.9 \\
    0870900301 & 2020-11-14 07:34:24 & 09:54:54.16 & +69:11:18.5 & 5.3 \\
    0870900101 & 2020-11-14 09:02:13 & 09:54:54.16 & +69:11:18.5 & 27.0 \\
    0904390101 & 2024-03-20 21:28:43 & 09:55:21.80 & +69:06:38.0 & 52.6 \\
    \hline
{\it NuSTAR} &&&&\\
    50401006002 & 2018-09-06 06:21:00 & 09:55:52.46	&   +69:08:28.31 & 8.8 \\
    50611001002 & 2020-11-12 03:41:00 & 09:54:38.47 &	+69:10:36.48 & 18.0 \\
    \hline
    {\it Chandra}/ACIS grating&&&&\\
6346  & 2005-07-14 01:43:51 & 09:55:33.20 & +69:03:55.10 & 55.2 \\
6347  & 2005-07-14 19:25:05 & 09:55:33.20 & +69:03:55.10 & 64.7 \\
6895  & 2006-04-24 08:17:47 & 09:55:33.20 & +69:03:55.10 & 14.7 \\
6896  & 2006-05-14 12:59:59 & 09:55:33.20 & +69:03:55.10 & 14.9 \\
6897  & 2006-06-09 18:12:58 & 09:55:33.20 & +69:03:55.10 & 14.9 \\
6898  & 2006-06-28 23:34:56 & 09:55:33.20 & +69:03:55.10 & 14.9 \\
6899  & 2006-07-13 13:40:29 & 09:55:33.20 & +69:03:55.10 & 15.1 \\
20624 & 2017-12-21 11:49:42 & 09:55:33.20 & +69:03:55.10 & 14.9 \\
\hline
\end{tabular}
    \caption{{\it XMM-Newton} and {\it NuSTAR} observations used in this work. The coordinates for the individual observations are the aim-points. }
    \label{tab:obs}
\end{table*}


Below, we use these data to examine the flare and non-flare X-ray emission.

\subsection{Flare Search}
In \citet{2026MNRAS.546f2237Z}, we analyzed all available {\it Chandra} Advanced CCD Imaging Spectrometer (ACIS) non-grating observations and detected 14 flares in M81F4 in 12 observations. 
Here we complement that analysis with available {\it Chandra}/ACIS grating observations and {\it XMM-Newton}/EPIC observations to further search for flares in M81F4, applying the Bayesian Blocks algorithm \citep{1998ApJ...504..405S} to the time sequence of the source counts, as detailed in \citet{2026MNRAS.546f2237Z}. Seven additional flares are identified in four {\it XMM-Newton} observations and summarized in Table~\ref{tab:flare}, but no flare is detected in any of the {\it Chandra}/ACIS grating observations. 
The individual light curves of all 21 flares, along with the Bayesian blocks, are presented in Appendix~\ref{app:individual}.

\subsection{Long-term Light Curve}
\label{sec:long}

To characterize the long-term X-ray behavior of M81F4, we combine the {\it Chandra}/ACIS non-grating and grating observations, {\it XMM-Newton}/EPIC observations, and {\it Swift}/XRT observations. All count rate measurements were restricted to the 0.5-8.0 keV band and converted to an unabsorbed 0.5-8.0 keV luminosity, assuming a certain spectral model (see below) and a nominal distance of 3.6 Mpc \citep{2001ApJ...553...47F}.

A total of 24 ACIS non-grating observations and 8 grating observations are utilized. 
The non-grating observations were reprocessed using CIAO v4.14 \citep{2006SPIE.6270E..1VF} and CALDB v4.9.8\footnote{\url{https://cxc.cfa.harvard.edu/ciao}}, following the standard procedures as in \citet{2026MNRAS.546f2237Z}. 
For the grating observations, we retained events satisfying \texttt{tg\_m=0} or \texttt{(tg\_m=99 \&\& tg\_smap<3)}, and further restricted to data registered by CCD 6 or CCD 7. This selection isolates the zeroth-order events, together with a small number of unassigned events close to the grating spatial region. For both ACIS grating and non-grating observations, source counts were extracted from within the 90\% enclosed counts radius (ECR), while background counts were extracted from a concentric annulus with inner and outer radii of 2 and 4 times the 90\% ECR, respectively. 
We used the CIAO task \texttt{aprates} to calculate the net source count rate and 1$\sigma$ error, correcting for the enclosed energy fraction. The count rate to flux conversion factor was derived from the best-fitted ACIS non-flare  spectral model (Section~\ref{subsec:spec}). 

A total of 8 {\it XMM-Newton} observations were utilized, for which we reduced the EPIC data using the {\it XMM-Newton} Science Analysis System (SAS; version 22.1.0)\citep{2004ASPC..314..759G}. 
The source counts were extracted from a circular region of $20^{\prime\prime}$ radius (approximately 75\% ECR) centered on M81F4, while the background counts were extracted from concentric annular regions with inner and outer radii of $40^{\prime\prime}$ and $60^{\prime\prime}$, respectively, on either the MOS1, MOS2 or PN detector.
We again used \texttt{aprates} to estimate the net source count rate and 1$\sigma$ error, correcting for the enclosed energy fraction. The count rate to flux conversion factor was derived from the best-fitted EPIC non-flare spectral model. 

Most individual {\it Swift}/XRT observations were too shallow to yield robust flux constraints. We therefore grouped the 486 observations into half-year bins and measured the source flux in each bin. Within each observation, source counts were extracted from a circular aperture centered on M81F4 with a radius of $\sim20^{\prime\prime}$, approximately corresponding to the 75\% ECR of the XRT point-spread function (PSF). Background counts were extracted from an annulus with inner and outer radii of $\sim40^{\prime\prime}$ and $\sim80^{\prime\prime}$, respectively. 
The source and background counts were then combined within each half-year bin, and the net source count rate and  1$\sigma$ error were estimated with \texttt{aprates}, again correcting for the enclosed energy fraction. The resulting net count rates were converted to 0.5-8.0 keV fluxes, using a conversion factor derived from the best-fit model of the {\it Chandra}/ACIS non-flare spectrum. 

In this way, the {\it Chandra}, {\it XMM-Newton}, and {\it Swift}/XRT measurements were placed on a common 0.5-8.0 keV luminosity scale and combined into a long-term light curve spanning about 25 years. 
\subsection{X-ray Spectra}
\label{sec:spec}
We extract source spectra from all available ACIS non-grating observations using the CIAO tool \texttt{specextract}. 
The source and background regions are the same as for constructing the long-term light curve.
For observations in which flares are detected, we divide the data into ``flare'' and ``non-flare'' intervals according to the Bayesian blocks, and extract the corresponding spectra separately. Observations without detected flares are assigned to the ``non-flare'' group. 
We then combine the individual spectra of the two groups using the CIAO tool \texttt{combine\_spectra}, with the corresponding ARFs and RMFs weighted by the effective exposure. 
The two combined spectra thus represent the average flare and non-flare spectral properties, respectively. 
M81F4 is located at a large off-axis angle in most ACIS observations, so the broadened {\it Chandra} PSF distributes the events over many detector pixels. We estimated the per-frame count rate within the brightest pixels and found it below the level at which photon pileup would significantly affect the photometric or spectral measurements. 
We do not include the ACIS grating observations for spectral analysis, because the instrumental response for the zeroth-order events at large off-axis angles is not well-tested and might cause unwanted systematics. 

Similarly, we extract the {\it XMM-Newton} EPIC-PN source and background spectra from the regions defined in Section~\ref{sec:long}. 
The MOS data were not included in the spectral analysis, because of their relatively low counting statistics. 
For observations containing flares, the PN events were separated into ``flare'' and ``non-flare'' intervals according to the Bayesian-blocks, while observations without detected flares were assigned to the ``non-flare'' group. The spectra and corresponding response files were generated using the standard SAS tasks and combined separately for the two groups using \texttt{epicspeccombine}. 

{\it Swift}/XRT spectra were extracted from the cleaned photon counting mode event files using \texttt{XSELECT}\footnote{\url{https://heasarc.gsfc.nasa.gov/docs/software/ftools/xselect/index.html}}.
Source and background spectra were extracted from the regions defined in Section~\ref{sec:long}. 
Since no flare can be identified in the {\it Swift}/XRT data, we extract a single set of spectra, which can potentially include unidentified flares. 
Ancillary response files were generated using \texttt{xrtmkarf} and the corresponding response matrices were obtained from the Swift CALDB. 

We utilized {\it NuSTAR} data to extend the spectral coverage of M81F4 to higher energies. Two {\it NuSTAR} observations, taken in 2018 and 2020, covered M81F4. The level 2 cleaned and calibrated event files, together with other necessary data products, were obtained from HEASARC. 
We extracted the FPMA and FPMB spectra with \texttt{nuproducts}, adopting a circular source region with a radius of 50\arcsec, and a background annulus with inner and outer radii of 70\arcsec and 100\arcsec, respectively. Several other {\it Chandra}-detected X-ray sources falling within this relatively large aperture are estimated to contribute less than 5\% of the total source counts and are therefore negligible. 

While the {\it Swift}/XRT and {\it NuSTAR} spectra may contain unresolved flare intervals, their fractional contribution to the total spectra is negligibly small (Section~\ref{subsec:spec}). 
The {\it Chandra}, {\it XMM-Newton}, and {\it Swift} spectra were analyzed over 0.5-8.0 keV, while the {\it NuSTAR} spectra were analyzed over 3-30 keV. Spectral fitting was performed with XSPEC v12.12.1. 

\section{Analysis and Results}
\label{sec:anal}
\subsection{Flare Properties}
\label{subsec:flare}

Table~\ref{tab:flare} summarizes the flares detected from M81F4 in the {\it Chandra} and {\it XMM-Newton} observations. For each flare, we list the start time, duration, peak count rate, reference count rate, peak luminosity, and integrated flare energy. The start time and duration are defined from the Bayesian blocks of flares. The reference count rate represents the local non-flare level used to estimate the excess flare emission. The peak luminosities and flare energies are reported in the 0.5-8.0 keV band, assuming the same distance and count-rate-to-flux conversion.

We detected 14 flares in 12 {\it Chandra} observations and 7 flares in 4 {\it XMM-Newton} observations, the first on MJD 53516 and the last on MJD 59167, demonstrating recurrent flaring activity over more than 15 years. Multiple flares are detected within individual observations, with the shortest observed separation being $\sim$6 ks (on MJD 53545). The flare occurrence rate also varies with time: 12 flares are detected in 178~ks in 2005 and 4 flares in 97~ks in 2011, corresponding to similar duty cycles of $\sim1\%$, whereas only 5 flares are found in 532~ks between 2016 and 2024, yielding a much lower duty cycle of $\sim0.2\%$. This decline appears to accompany the decrease in the source flux beginning around 2018 (Section~\ref{subsec:lc}).

Flares are not found in all observations. In particular, no flare is found in any of the ACIS grating observations. 
Our simulations of light curves comprising a constant baseline and artificial flares (see Appendix~\ref{app:completeness} for a detailed treatment of flare detection completeness) indicate that the absence of detected flares in the ACIS grating observations is due to their relatively low effective area and thus substantially reduced sensitivity. In contrast, the last {\it XMM-Newton} observation (ObsID 0904390101), with an exposure of 52.6~ks, was sufficiently sensitive if flares occurred during the observation. Therefore, the lack of flare detection in this observation might be intrinsic and related to the drop of the mean source luminosity since 2022 (see Section~\ref{subsec:lc}).
%

Considering all flares detected during the available {\it Chandra}/ACIS non-grating exposures, the long-term mean recurrence time is $\sim0.5$ day, based on 625.6 ks of total effective exposure and 14 detected flares. An independent estimate from the {\it XMM-Newton} observations gives a similar recurrence time of $\sim0.5$ day from 315.6 ks of exposure and 7 flares. Depending on the observing epoch, incomplete flare detection may cause the {\it Chandra} recurrence times to be overestimated by $\sim20\%$, whereas the incompleteness correction to the {\it XMM-Newton} recurrence time estimate is negligible (Appendix~\ref{app:completeness}).

\begin{table*}
\centering
\begin{tabular}{ccccccccc}
\hline
Seq & ObsID & Instrument & start time & Duration & $CR_{\rm peak}$ & $CR_{\rm ref}$ & $L_{\rm peak}$ & $E_{\rm flare}$ \\
 & & & MJD & sec & count s$^{-1}$ & count s$^{-1}$ & $10^{39}$ erg s$^{-1}$ & $10^{41}$ erg \\
(1) &(2) &(3)&(4)&(5)&(6)&(7)&(8)&(9)\\
\hline
{\it Chandra} & & & & & & & & \\
1 & 5935 &ACIS-S& 53516.1894 & 94 & 0.16$\pm$0.09 & 0.022$\pm$0.001 & 2$\pm$1 & 1.6$\pm$0.6 \\
2 & 5936 &ACIS-S& 53518.8562 & 162.25 & 0.22$\pm$0.08 & 0.019$\pm$0.002 & 4$\pm$2 & 4$\pm$1 \\
3 & 5936 &ACIS-S& 53518.9470 & 81.12 & 0.4$\pm$0.2 & 0.019$\pm$0.002 & 7$\pm$3 & 3$\pm$0.8 \\
4 & 5937 &ACIS-S& 53522.4696 & 217.37 & 0.14$\pm$0.06 & 0.013$\pm$0.001 & 4$\pm$1 & 3$\pm$1 \\
5 & 5938 &ACIS-S& 53525.0053 & 97.24 & 0.4$\pm$0.1 & 0.014$\pm$0.001 & 9$\pm$3 & 4$\pm$1 \\
6 & 5941 &ACIS-S& 53533.0073 & 181.53 & 0.25$\pm$0.08 & 0.017$\pm$0.001 & 6$\pm$2 & 5$\pm$1 \\
7 & 5942 &ACIS-S& 53536.0785 & 131.37 & 0.3$\pm$0.1 & 0.02$\pm$0.001 & 5$\pm$2 & 2.9$\pm$0.9 \\
8 & 5943 &ACIS-S& 53539.6177 & 82.66 & 0.3$\pm$0.1 & 0.024$\pm$0.001 & 5$\pm$2 & 1.6$\pm$0.6 \\
9 & 5945 &ACIS-S& 53545.3099 & 106.97 & 0.5$\pm$0.1 & 0.026$\pm$0.002 & 7$\pm$2 & 2.9$\pm$0.7 \\
10 & 5945 &ACIS-S& 53545.3744 & 240 & 0.17$\pm$0.06 & 0.026$\pm$0.002 & 3$\pm$1 & 2.8$\pm$0.8 \\
11 & 5948 &ACIS-S& 53554.1537 & 191.24 & 0.24$\pm$0.08 & 0.027$\pm$0.002 & 4$\pm$1 & 2.9$\pm$0.8 \\
12 & 5949 &ACIS-S& 53557.4171 & 213.92 & 0.21$\pm$0.07 & 0.03$\pm$0.002 & 3$\pm$1 & 2.4$\pm$0.8 \\
17 & 19981 &ACIS-I& 57764.6018 & 127.21 & 0.16$\pm$0.08 & 0.0121$\pm$0.0007 & 5$\pm$3 & 3$\pm$1 \\
18 & 18050 &ACIS-I& 57778.1380 & 329.85 & 0.11$\pm$0.04 & 0.0131$\pm$0.0006 & 3$\pm$1 & 5$\pm$1 \\
\hline
{\it XMM-Newton} & & & & & & & & \\
13 & 0657802001 & PN+MOS & 55644.6840 & 37.5 & $1.2 \pm 0.4$ & $0.112 \pm 0.009$ & $5 \pm 2$ & $1.1 \pm 0.2$ \\
14 & 0657801801 & MOS & 55830.2538 & 178.4 & $0.22 \pm 0.08$ & $0.021 \pm 0.001$ & $2.2 \pm 0.8$ & $2.9 \pm 0.5$ \\
15 & 0657802201 & PN+MOS & 55888.1157 & 144.7 & $0.3 \pm 0.1$ & $0.044 \pm 0.002$ & $1.3 \pm 0.4$ & $1.4 \pm 0.2$ \\
16 & 0657802201 & PN & 55888.2028 & 62.3 & $0.30\pm0.2$ & $0.03\pm0.004$ & $2\pm1$ & $1.2\pm0.6$ \\
19 & 0870900101 & PN+MOS & 59167.5035 & 140.5 & $0.3 \pm 0.1$ & $0.058 \pm 0.002$ & $1.1 \pm 0.4$ & $1.2 \pm 0.2$ \\
20 & 0870900101 & PN+MOS & 59167.5945 & 178.7 & $0.4 \pm 0.1$ & $0.059 \pm 0.002$ & $1.5 \pm 0.4$ & $2.1 \pm 0.3$ \\
21 & 0870900101 & PN+MOS & 59167.6714 & 164.3 & $0.4 \pm 0.1$ & $0.059 \pm 0.003$ & $1.5 \pm 0.4$ & $1.7 \pm 0.3$ \\
\hline
\end{tabular}
\caption{Information of flares detected in {\it Chandra} and {\it XMM-Newton} observations. 
Column (1) gives the flare sequence number; Columns (2) and (3) list the observation identifier and detector(s), respectively.
The flare start time and duration in Columns (4) and (5) are defined by the Bayesian-block interval associated with each flare.
$CR_{\rm peak}$ in Column (6) is the peak source count rate, estimated by dividing the flare interval into five equal time bins and adopting the maximum count rate among them.
$CR_{\rm ref}$ in Column (7) is the local non-flare count rate used as the reference level for estimating the excess flare emission.
$L_{\rm peak}$ in Column (8) is the 0.5-8.0~keV peak luminosity converted from $CR_{\rm peak}$ using the count-rate-to-flux conversion appropriate for the corresponding observation and assuming a distance of 3.6~Mpc.
$E_{\rm flare}$ in Column (9) is the 0.5-8.0~keV energy released in excess of the reference level over the flare interval.
All quoted uncertainties are at the $1\sigma$ confidence level.}
\label{tab:flare}
\end{table*}

As shown in Appendix~\ref{app:individual}, some individual light curves are qualitatively consistent with a fast rise followed by a slower decay, 
although the limited photon statistics preclude a detailed characterization of their temporal profiles. Considering the fact that the durations and peak luminosities of the detected flares are broadly consistent with each other (Table~\ref{tab:flare}), we construct an average flare profile by stacking the flares detected in the {\it Chandra} and {\it XMM-Newton} observations, to better constrain their characteristic temporal properties. For the {\it XMM-Newton} observations, only the PN events were used to construct the stacked profile.

For each flare, we divide the Bayesian flare block into five equal-duration segment and selected the segment with the highest 0.5--8.0 keV count rate as the estimated peak time. If multiple segments had the same maximum count rate, the earliest was selected. We aligned the photon arrival times relative to these peak estimates and constructed individual flare profile over -500 to 1000 s using 40 s bins. We then averaged the response-corrected individual profiles in each bin to obtain the stacked 0.5--8.0 keV absorbed photon flux profile, with the statistical uncertainty obtained by propagating the Poisson counting error of individual flares. 

The hardness ratio, defined as ${\rm HR}=(H-S)/(H+S)$ with $H$ for the $2.0-8.0$ keV band and $S$ for the $0.5-2.0$ keV band, shows no statistically significant evolution during the flare. In particular, no progressive spectral softening is detected during the decay phase, in contrast to the cooling tails commonly observed in classical Type I X-ray bursts \citep{2021ASSL..461..209G}.

The resulting averaged profiles, plotted in Figure~\ref{fig:profile}, preserve the fast rise slow decay trend suggested by several individual flares (see Appendix~\ref{app:individual}). 
We characterize the average flare profiles with a Norris Fast-Rise Exponential-Decay (FRED) model \citep{Norris_2005} 
\begin{equation}
R(t)=
\left\{
\begin{array}{ll}
C + A\lambda \exp\left[-\displaystyle\frac{\tau_1}{t-t_s}
-\displaystyle\frac{t-t_s}{\tau_2}\right], & t>t_s, \\\\[6pt]
C, & t\leq t_s .
\end{array}
\right.
\end{equation}
Here $C$ is the baseline photon flux, $A$ is the peak excess photon flux above the baseline, $t_s$ is the onset of the modeled pulse. The positive parameters $\tau_1$ and $\tau_2$ control the rising and decaying timescale, and $\lambda=\exp\left(2\sqrt{\frac{\tau_1}{\tau_2}}\right)$ is a normalization factor. The model peaks at $t_{\rm pk}=t_s+\sqrt{\tau_1\tau_2}$. 

We fit the stacked profiles independently for \textit{Chandra} and \textit{XMM-Newton}, using weighted nonlinear least squares, with each time bin weighted by the inverse variance of its photon flux. 
We define $t_{\rm rise}$ and $t_{\rm decay}$ as the time intervals from the rising point at $e^{-3}$ of the peak excess above the baseline to the peak, and from the peak to the declining point at the same level, respectively. For the FRED model, 
\begin{align}
    t_{\rm rise}=\frac{\tau_2}{2}(\sqrt{9+12\sqrt{\tau_1/\tau_2}}-3),\\
    t_{\rm decay}=\frac{\tau_2}{2}(\sqrt{9+12\sqrt{\tau_1/\tau_2}}+3).
\end{align} 

The fits yields $t_{\rm rise}=97\pm16~\mathrm{s}$ and $t_{\rm decay}=201\pm29~\mathrm{s}$ for \textit{Chandra}, and $t_{\rm rise}=143\pm37~\mathrm{s}$ and $t_{\rm decay}=293\pm64~\mathrm{s}$ for \textit{XMM-Newton}.
The average peak luminosities, calculated from the best-fit values of $C$ and $A$, are $3.4\pm0.3\times 10^{39}~\rm erg~s^{-1}$ and $1.8\pm0.2\times 10^{39}~\rm erg~s^{-1}$ for {\it Chandra} and {\it XMM-Newton}, respectively. The quoted uncertainties are $1\sigma$ errors propagated from the fit covariance matrix, including the covariance between $C$ and $A$. 

\begin{figure*}
    \centering
    \includegraphics[width=0.7\linewidth]{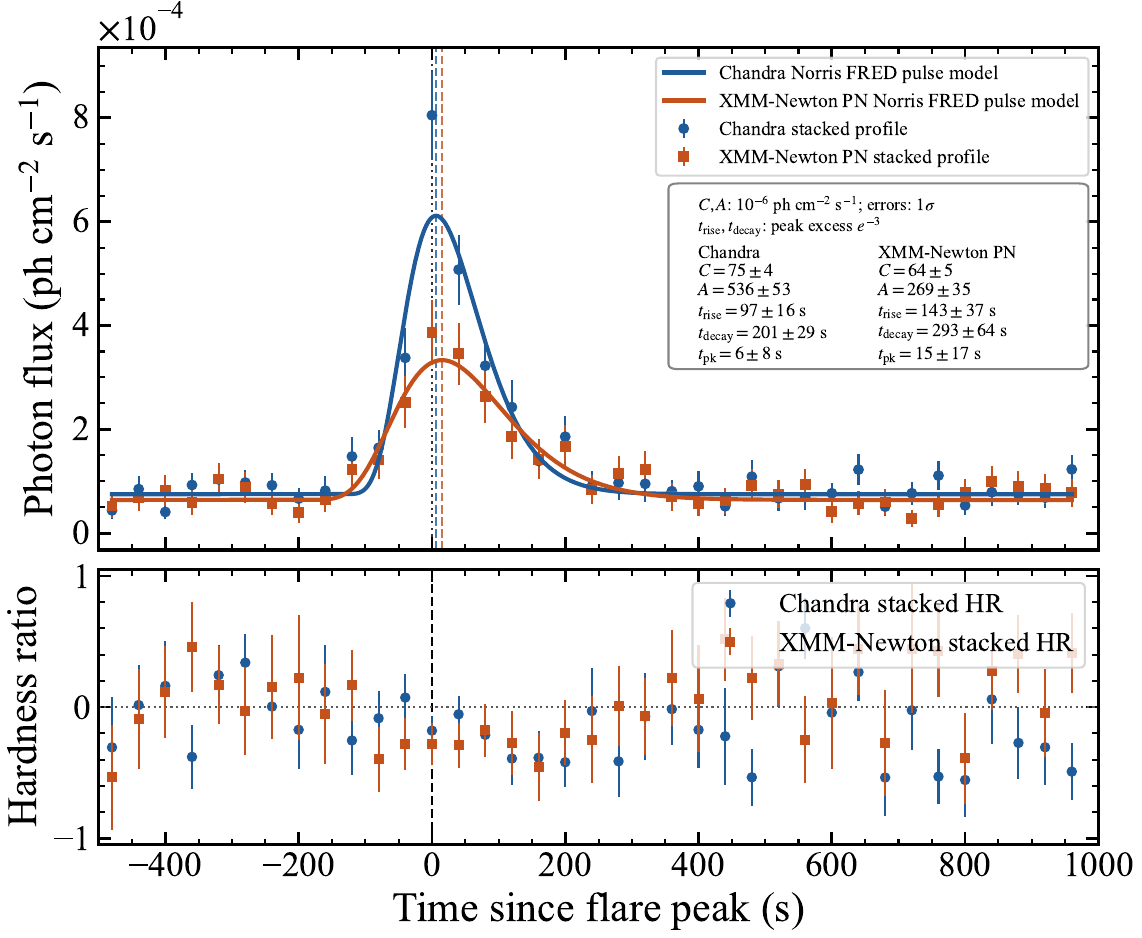}
    \caption{{\it Upper panel}: Stacked flare profile of M81F4, constructed by aligning individual flares at their estimated peak times and averaging their response-corrected profiles in uniform 40 s bins. For each flare, $t=0$ corresponds to the midpoint of the peak segment among five equal-duration segments of its flare Bayesian block. 
    The {\it Chandra} and {\it XMM-Newton} flare profiles are shown by blue circles and red squares, respectively. The FRED model is shown by the solid curves, with best-fitted parameters denoted in the inset. The rise and decay times are measured between the model peak and the points where the excess above baseline falls to $e^{-3}$ of its peak value. 
    {\it Lower panel:} Corresponding hardness ratio as a function of time, showing no significant spectral evolution during the flare.}
    \label{fig:profile}
\end{figure*}

\subsection{Long-term Variability}
\label{subsec:lc}
\begin{figure*}
    \centering
    \includegraphics[width=0.99\linewidth]{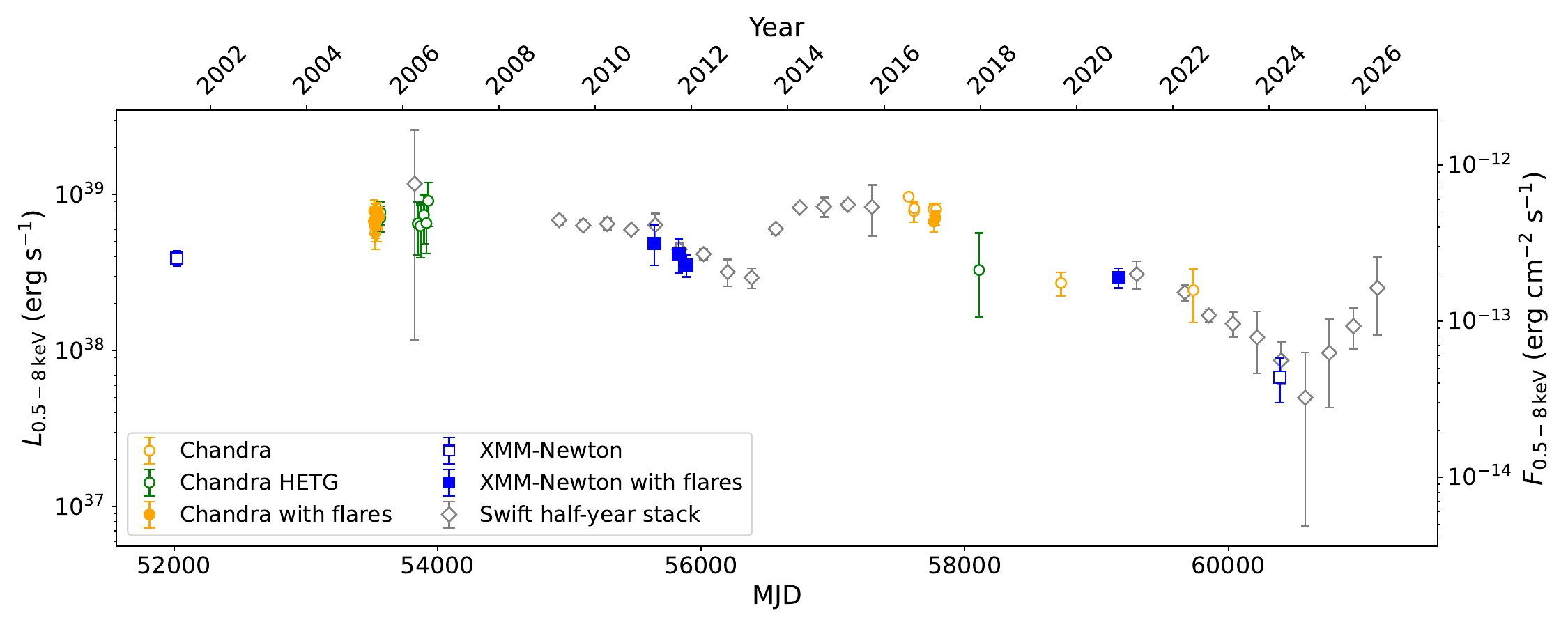}
    \caption{Long-term variation of the 0.5-8.0 keV luminosity  of M81F4. {\it Chandra}, {\it XMM-Newton}, and {\it Swift}/XRT data are shown as circles, squares, and diamonds, respectively. The {\it Swift}/XRT data are grouped into half-year bins to improve photon statistics. Filled symbols mark observations in which short flares are detected, while open symbols indicate observations without detected flares. The occurrence of flares  spanning nearly two decades indicates that the flaring activity is persistent, although the intrinsic occurrence rate may have varied. }
    \label{fig:long}
\end{figure*}

As described in Section~\ref{sec:data}, we constructed the long-term light curve of M81F4 by combining the available {\it Chandra}, {\it XMM-Newton}, and {\it Swift}/XRT observations, as shown in Figure~\ref{fig:long}. Over two decades, the source has remained persistently luminous, with an unabsorbed 0.5-8.0 keV luminosity in the range of $\sim 10^{38}-10^{39}~\rm erg~s^{-1}$. Although variability is present on both short- and long-timescales, the source never drops to a truly quiescent state. This persistently high luminosity indicates that the short flares occur in an actively accreting system, rather than arising from a quiescent or transient source that would have appeared faint for most of the observing time. 

Notably, the long-term light curve exhibits flux decreases around 2013 and 2024 (Figure~\ref{fig:long}), 
most evident in the {\it Swift}/XRT half-year bins and supported by {\it Chandra} and {\it XMM-Newton} data. 
The flux decreases are separated by $\sim11$ yr, but sparse sampling and the small number preclude establishing a coherent period. Similar variations in persistent Galactic XRBs have been attributed to changes in donor mass transfer or third-body modulation \citep{2010MNRAS.401..355D}.

\subsection{Spectral properties}
\label{subsec:spec}
Because the source flux has varied substantially (Section~\ref{subsec:lc}) and the spectra acquired by different missions were not simultaneous, we fit them separately. No significant spectral variation is found among observations taken with the same instrument, therefore we focus on the combined spectra with improved S/N. The spectral fitting results are summarized in Table~\ref{tab:spec}.

\begin{figure*}
    \centering
    \begin{tikzpicture}
        \node[inner sep=0pt] (main) at (0,0) {
            \includegraphics[width=0.95\linewidth]{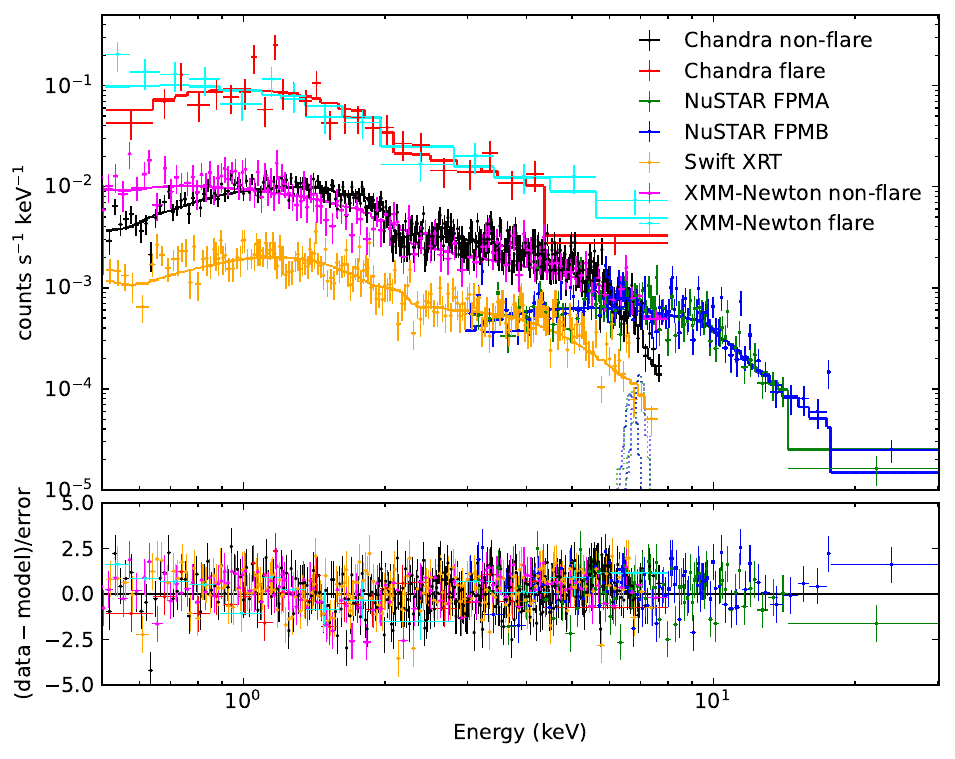}
        };
        \begin{scope}[
            x={($(main.south east)-(main.south west)$)},
            y={($(main.north west)-(main.south west)$)},
            shift={(main.south west)}
        ]

            \node[anchor=south west, inner sep=0pt] at (0.13,0.36) {
                \includegraphics[width=0.25\linewidth]{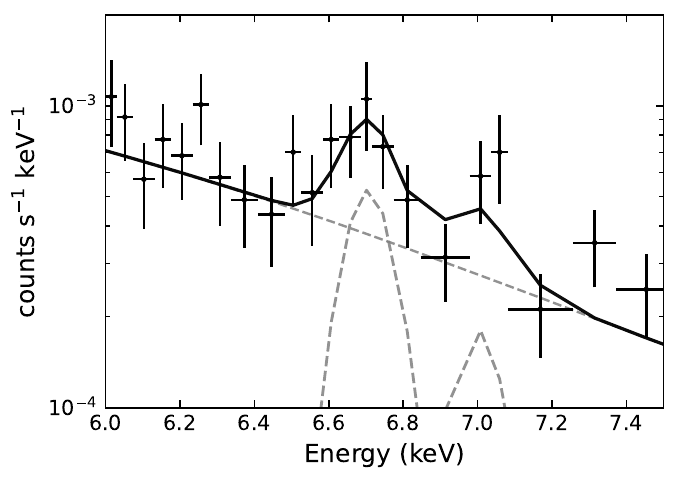}
            };
        \end{scope}
    \end{tikzpicture}

\caption{
X-ray spectra of M81F4 obtained with \textit{Chandra}, \textit{NuSTAR},
\textit{Swift}/XRT, and \textit{XMM-Newton}.
The solid curves show the best-fitting spectral models. 
The \textit{Chandra} and \textit{XMM-Newton} non-flare spectra are fitted with an absorbed powerlaw plus two Gaussian components, while the \textit{NuSTAR} spectrum is fitted with an absorbed broken powerlaw plus two Gaussian components. 
The \textit{Chandra} and \textit{XMM-Newton} flare spectra and the \textit{Swift}/XRT spectrum are fitted with absorbed power-law models. 
The corresponding residuals are shown in the lower panels. 
The inset highlights the 6.0--7.5~keV region of the \textit{Chandra} non-flare spectrum, showing an emission feature near 6.7~keV consistent with Fe~XXV K$\alpha$, and a possible weaker feature near 7.0~keV
consistent with Fe~XXVI Ly$\alpha$.
}
\label{fig:spec}
\end{figure*}

The {\it Chandra} spectra provide the best counting statistics. As shown in Figure~\ref{fig:spec}, the non-flare spectrum is dominated by a hard continuum. An absorbed \texttt{powerlaw+gaussian+gaussian} model gives a photon index of $\Gamma=1.15^{+0.04}_{-0.03}$ (90\% confidence level) with the absorption column density fixed at the Galactic foreground value of $N_{\rm H}=3\times 10^{20}~\rm cm^{-2}$, taken from HI4PI survey \citep{2016A&A...594A.116H}. Adding a blackbody component significantly improves the fit from $\chi^2$/dof $341.92/323$ to $317.57/321$, yielding $kT=0.26\pm0.03$~keV and $L_{\rm bb}=4.0_{-1.4}^{+1.5}\times10^{37}~\rm erg~s^{-1}$, corresponding to $\sim6\%$ of the total unabsorbed 0.5-8.0 keV luminosity of $L_{0.5-8.0}=6.2_{-0.2}^{+0.2}\times10^{38}~\rm erg~s^{-1}$. The power-law continuum correspondingly becomes harder, with $\Gamma=0.9_{-0.1}^{+0.1}$.

The non-flare spectrum shows two line-like excesses near 6.7 and 7.0 keV (see inset in Figure~\ref{fig:spec}), plausibly associated with Fe XXV K$\alpha$ and Fe XXVI Ly$\alpha$, respectively. 
We assessed their significance through Monte Carlo simulations using XSPEC \texttt{simftest} \citep{2002ApJ...571..545P}. For each feature, we generate 10000 sets of simulated spectra, obtaining $p$-vlaue of 0.0074 and 0.0897 for the 6.7 and 7.0 keV features, respectively. These correspond to one-sided Gaussian-equivalent significance of $2.5\sigma$ and $1.3\sigma$, respectively. The fitted equivalent widths are $0.18^{+0.10}_{-0.08}$ keV and $0.10^{+0.08}_{-0.08}$ keV, respectively.

In contrast, the flare spectrum is featureless and is well described by an absorbed \texttt{powerlaw} with $\Gamma=1.4^{+0.1}_{-0.2}$, slightly softer than the non-flare spectrum, and an unabsorbed 0.5-8.0 keV luminosity of $L_{0.5-8.0}=2.8_{-0.3}^{+0.3}\times 10^{39}~\rm erg~s^{-1}$. Adding a \texttt{bbody} component produces no significant improvement ($\Delta\chi^2=0.01$), and its temperature cannot be constrained. Fixing the temperature at $0.26$ keV gives a 3$\sigma$ upper limit of $L_{\rm bb}<4\times10^{38}~\rm erg~s^{-1}$. Thus, the flare emission remains dominated by a hard power-law continuum rather than a thermal component.

The {\it XMM-Newton}/PN spectra show the same overall behavior, with $\Gamma=1.14^{+0.08}_{-0.08}$ in the non-flare state and $\Gamma=1.4^{+0.3}_{-0.3}$ during flares. A weak soft component is again allowed in the non-flare spectrum, with $kT=0.22_{-0.04}^{+0.04}~\rm keV$ and a contribution of $<10\%$ to the 0.5-8.0 keV luminosity. The {\it Swift}/XRT spectrum is also hard, with $\Gamma=1.07^{+0.06}_{-0.06}$.

At higher energies, the {\it NuSTAR} spectra show clear spectral curvature and are better described by a broken power-law model. The best-fit break energy is $9.6^{+1.0}_{-0.8}$ keV, with photon index of $1.1^{+0.2}_{-0.2}$ and $2.8^{+0.6}_{-0.4}$ below and above the break, respectively. 
Extrapolating the best-fit model to 0.5~keV gives an unabsorbed 0.5--30 keV luminosity of $L_{0.5-30~\rm keV} =8.7_{-0.6}^{+0.7}\times10^{38}\rm ~erg~s^{-1}$, roughly five times the Eddington luminosity of a canonical $1.4~M_\odot$ NS.

\begin{sidewaystable*}
\centering
\fontsize{8pt}{15pt}\selectfont 
\setlength{\tabcolsep}{1pt}
\begin{tabular}{lllccccccc}
\hline
\hline
\tlcell{2.2cm}{Instrument} &
\tlcell{3.5cm}{Model} &
\tlcell{1.2cm}{$\Gamma$} &
\tcell{2.0cm}{$N_{\rm PL}$} &
\tcell{1.8cm}{$kT_{\rm bb}$} &
\tcell{2cm}{$N_{\rm bb}$} &
\tcell{1.8cm}{$F_{6.7\,\rm keV}$} &
\tcell{1.8cm}{$F_{7.0\,\rm keV}$} &
\tcell{1.8cm}{$L_{0.5-8.0}$} &
\tcell{1.5cm}{$\chi^2/\mathrm{dof}$}
\\
&
&
& $\mathrm{ph~keV^{-1}~cm^{-2}~s^{-1}} $
& keV
&
& $\mathrm{ph~cm^{-2}~s^{-1}} $
& $\mathrm{ph~cm^{-2}~s^{-1}} $
& $ 10^{38}~\mathrm{erg~s^{-1}} $ 
& \\
(1)
&(2)
&(3)
&(4)
&(5)
&(6)
&(7)
&(8)
&(9)
&(10)
\\
\hline
\parbox[c]{1.5cm}{\textit{Chandra}\\[-5pt]
non-flare}
& 
\parbox[c]{2.0cm}{\raggedright\ttfamily
tbabs*(powerlaw+\\[-5pt]
gaussian+gaussian)}
& $1.15_{-0.03}^{+0.04}$
& $3.7_{-0.1}^{+0.1}\times 10^{-5}$
& --
& --
& $10_{-5}^{+5}\times 10^{-7}$
& $6_{-4}^{+5}\times 10^{-7}$
& $5.9_{-0.1}^{+0.1}$ 
& 341.92/323\\
[+8pt]
\parbox[c]{1.5cm}{\textit{Chandra}\\[-5pt]
non-flare}
& 
\parbox[c]{2.0cm}{\raggedright\ttfamily
tbabs*(powerlaw+bbody+\\
gaussian+gaussian)}
& $0.9_{-0.1}^{+0.1}$
& $2.6_{-0.4}^{+0.4}\times 10^{-5}$
& $0.26_{-0.03}^{+0.03}$
& $3.6_{-1.2}^{+1.2}\times10^{-7}$
& $8_{-5}^{+5}\times 10^{-7}$
& $<1.2\times 10^{-6}$
& $6.2_{-0.2}^{+0.2}$
& 317.57/321\\
[+8pt]
\parbox[c]{1.5cm}{\textit{Chandra}\\[-5pt]
flare}
& \texttt{tbabs*powerlaw}
& $1.4_{-0.2}^{+0.1}$
& $2.3_{-0.3}^{+0.3}\times 10^{-4}$
& --
& --
& --
& --
& $28_{-3}^{+3}$ 
& 21.97/25\\
[+8pt]
\parbox[c]{1.5cm}{\textit{Chandra}\\[-5pt]
flare}
& 
\parbox[c]{2.0cm}{\raggedright\ttfamily
tbabs*(powerlaw\\[-5pt]+bbody)}
& $1.3_{-0.5}^{+0.2}$
& $2.1_{-1.0}^{+0.7}\times 10^{-4}$
& $^a$$0.26$
& $<6\times10^{-6}$
& --
& --
& $29_{-4}^{+4}$
& 21.96/24\\
\textit{XMM-Newton} flare
& \texttt{tbabs*powerlaw}
& $1.4_{-0.3}^{+0.3}$
& $1.6_{-0.4}^{+0.3}\times10^{-4}$
& --
& --
& --
& --
& $20_{-4}^{+5}$
& 11.80/11 \\
\textit{XMM-Newton} non-flare
& \parbox[c]{2.0cm}{\raggedright\ttfamily
tbabs*(powerlaw+\\
gaussian+gaussian)}
& $1.14_{-0.08}^{+0.08}$
& $1.6_{-0.1}^{+0.1}\times10^{-5}$
& --
& --
& $<7\times10^{-7}$
& $<9\times 10^{-7}$
& $2.5_{-0.2}^{+0.2}$
& 98.65/110 \\
[+8pt]
\textit{XMM-Newton} non-flare
& 
\parbox[c]{2.0cm}{\raggedright\ttfamily
tbabs*(powerlaw+bbody+\\
gaussian+gaussian)}
& $0.8_{-0.2}^{+0.2}$
& $1.0_{-0.2}^{+0.3}\times 10^{-5}$
& $0.22_{-0.04}^{+0.04}$
& $1.8_{-0.8}^{+0.8}\times10^{-7}$
& $<6\times 10^{-7}$
& $<7\times 10^{-7}$
& $2.6_{-0.2}^{+0.2}$
& 85.10/108\\
\textit{Swift}/XRT
& \texttt{tbabs*powerlaw}
& $1.07_{-0.06}^{+0.06}$
& $3.3_{-0.2}^{+0.2}\times 10^{-5}$
& --
& --
& --
& --
& $5.6_{-0.2}^{+0.3}$ 
& 198.15/185\\
[+8pt]
\textit{NuSTAR}
& 
\parbox[c]{2.0cm}{\raggedright\ttfamily
tbabs*(powerlaw+\\
gaussian+gaussian)}
& $1.54_{-0.08}^{+0.08}$
& $6_{-1}^{+1}\times 10^{-5}$
& --
& --
& $<1.1\times 10^{-6}$
& $<1.2\times 10^{-6}$
& -- 
& 225.54/149\\
[+8pt]
\textit{NuSTAR}
& 
\parbox[c]{2.0cm}{\raggedright\ttfamily
tbabs*(bknpower+\\
gaussian+gaussian)}
& \parbox[c]{2.3cm}{$^b$$1.1_{-0.2}^{+0.2}$\\
$^b$$2.8_{-0.4}^{+0.6}$}
& $2.7_{-0.8}^{+1.0}\times 10^{-5}$
& $^c$$9.6_{-0.8}^{+1.0}$
& --
& $<9.7\times10^{-7}$
& $<1.0\times 10^{-6}$
& --
&179.43/147
\\[+8pt]
\hline
\end{tabular}
\caption{Spectral fitting results of M81F4. The absorption column density was fixed at $N_{\rm H}=3\times10^{20}~{\rm cm^{-2}}$ in all fits, because there was no evidence for significant additional absorption in the spectra.  
The centroid energy of the lines is fixed at 6.7 and 7.0 keV, respectively, and the Gaussian widths were fixed at zero. The Gaussian normalizations, $\mathrm{F}_{6.7\rm~keV}$ and $\mathrm{F}_{7.0\rm~keV}$, are the total line photon fluxes (or upper limits) in units of $\mathrm{photons~cm^{-2}~s^{-1}}$. 
The quoted total luminosities are unabsorbed values in the 0.5--8.0~keV band, calculated assuming a distance of 3.6~Mpc. 
All parameter uncertainties are quoted at the 90\% confidence level for one parameter of interest, while upper limits are given at the $3\sigma$ confidence level. 
$^a$The blackbody temperature of the {\it Chandra} flare spectrum is fixed to the same value as the non-flare spectrum.
$^b$The photon indices below and above the break energy of the broken power-law.
$^c$The break energy of the broken power-law.
}
\label{tab:spec}
\end{sidewaystable*}

\subsection{Potential Optical and Radio Counterparts}
\label{sec:optical}
We searched archival {\it HST}/ACS/WFC images in the F435W, F606W, and F814W bands for possible optical counterparts of M81F4. Several bright, point-like sources lie within the 90\% positional uncertainty ($1^{\prime\prime}$) allowed by the {\it Chandra} absolute astrometry (see Appendix~\ref{app:optical} for details of the {\it HST} images, astrometry, and photometry). 
Assuming only Galactic foreground extinction, the brightest source within the $1''$ Chandra error circle has $M_{\rm F435W}\simeq-3.19$. 
Comparison with the MIST model rules out luminous OB supergiant or main-sequence star earlier than B0V ($\sim15~M_\odot$; \citealp{2006MNRAS.371..185W}) in the vicinity of M81F4. 
On the other hand, the photometry of several sources (e.g., \#5, \#12) is consistent with a subgiant or red giant whose initial mass is about $4-5\rm~M_\odot$. 
One of these sources could a plausible counterpart of M81F4 (Figure~\ref{fig:hst}), serving as the donor to an accreting compact object (see further discussions in Section~\ref{subsec:NS}).
However, we caution that the optical emission of the donor star in an XRB might be contaminated by the accretion disk, hence the three-band photometry does not allow a unique donor classification. 
Moreover, a fainter donor below the {\it HST} detection limit, such as a white dwarf or low-mass main-sequence star, cannot be ruled out.

No strong radio counterpart of M81F4 is found in the available high-resolution VLA images. A 6 GHz observation reported by \cite{2019MNRAS.489.1181A} has an rms noise of 8 ${\mu}$Jy near the source position, corresponding to a $5\sigma$ upper limit of monochromatic luminosity $L_{\rm 6 GHz}<3.7\times10^{33}~{\rm erg~s^{-1}}$ at the distance of M81. This limit is not sufficiently deep to be constraining,
since BH XRBs can have radio luminosities of only $\sim10^{30}$--$10^{31}~\rm erg~s^{-1}$ at $L_{\rm X}\sim10^{38}~\rm erg~s^{-1}$, while NS XRBs are on-average fainter at comparable X-ray luminosities \citep{2006MNRAS.366...79M}.
 
\section{Discussion}
The analyses presented in Section~\ref{sec:anal} establish several unusual properties of M81F4. The source has remained persistently bright for more than two decades, exhibits recurrent flares with a mean recurrence time of $\sim0.5$ day and separations as short as 6 ks, and maintains a hard X-ray continuum extending to at least $\sim$10 keV. 

A background active galactic nucleus (AGN) or foreground stellar source is strongly disfavored. The tentative $\sim6.7$ and $\sim7.0$ keV features are consistent with Fe XXV K$\alpha$ and Fe XXVI Ly$\alpha$ of negligible redshift, arguing against a distant AGN. A nearby flaring low-mass star is unlikely given the unusually hard X-ray spectrum and the absence of a bright optical counterpart in either the {\it HST} images or the Gaia DR2 catalog\footnote{\url{https://www.cosmos.esa.int/web/gaia/dr2}}. A foreground massive star is similarly disfavored because its wind-shock emission would generally be dominated by a much softer thermal spectrum. If located at 10 kpc, M81F4 would have $L_{\rm X}\sim10^{33}~{\rm erg~s^{-1}}$, comparable to cataclysmic variables (CVs) or quiescent XRBs, neither of which provide a natural analogue for the observed recurrent minute-scale flares or hard spectrum.

We therefore favor a physical association of M81F4 with M81, which implies a persistent 0.5-8 keV luminosity of $\sim10^{38}-10^{39}\rm~erg~s^{-1}$ and flare peaks reaching a few $10^{39}\rm~erg~s^{-1}$. Below we discuss the origin of these luminous flares and the nature of the underlying system.

\subsection{Ruling Out Classical X-ray Bursts}

Type I X-ray bursts are generally understood as unstable thermonuclear burning on the surface of an accreting NS \citep{2008ApJS..179..360G}. Classical Type I bursts typically show rapid spectral evolution and a cooling tail, whereas no significant spectral cooling is detected during the M81F4 flares. Moreover, the flares have typical radiated energies of $\sim10^{41}~{\rm erg}$, substantially exceeding the $10^{39}$--$10^{40}~{\rm erg}$ range of classical Type I bursts. Rare thermonuclear superbursts can reach comparable energies \citep{2023MNRAS.521.3608A}, 
but repeatedly accumulating sufficient nuclear fuel on the NS surface to power $\sim10^{41}$ erg events on a $\sim0.5$ day cadence would be very challenging. Taken together, these discrepancies make a thermonuclear burst origin unlikely.

Type II X-ray bursts, on the other hand, usually lack the characteristic cooling tails of thermonuclear Type I bursts and are generally attributed to instabilities in the accretion flow. 
Nevertheless, M81F4 is phenomenologically distinct from the classical Type II bursters Rapid Burster and GRO J1744-28 (both consisting of an NS accreting from a low-mass donor), whose bursts typically recur on timescales of seconds to minutes and often show complex, flat-topped, or multi-peaked profiles \citep{2015MNRAS.449..268B,2018MNRAS.481.2273C}. The flares of M81F4 instead show a relatively simple fast-rise, slow-decay profile and recur on a much longer timescale of at least hours. Thus, while an accretion-instability origin remains possible, M81F4 does not simply mirror the known Type II bursters.

M81F4 is also distinct from the ultraluminous X-ray bursts reported by \citet{2016Natur.538..356I}, which occurred in two globular clusters or ultra-compact dwarfs (i.e., old stellar environments) around the elliptical galaxies NGC 4636 and Cen A. Although those events show similarly fast-rise, slow-decay profiles, they typically recur less frequently, last $\sim1-2$ ks, and reach peak luminosities ($1-9\times10^{40}~{\rm erg~s^{-1}}$) nearly two orders of magnitude above the pre-flare level, compared with only a factor of $\sim5-30$ (with substantial statistical uncertainty) in M81F4. The differences in recurrence, duration, flare amplitude, and host environment argue against M81F4 being a direct analog of the ultraluminous X-ray bursts.

\subsection{Comparison with Bursting HMXBs}
High-mass X-ray binaries (HMXBs), in which a massive donor feeds the compact object with it mass-losing winds, often exhibit strong, short-term X-ray variability. 
In particular, supergiant fast X-ray transients (SFXTs) typically spend most of their time at $10^{32}-10^{34}~{\rm erg~s^{-1}}$ and flare to $10^{36}-10^{37}{\rm erg~s^{-1}}$, reaching dynamic ranges of $10^3-10^5$, which are thought to be powered by episodic wind accretion \citep{2017SSRv..212...59M}. M81F4 instead remains persistently luminous at $10^{38}-10^{39}~{\rm erg~s^{-1}}$, with flares representing only modest enhancements above this bright state. A Be-HMXB interpretation is likewise disfavored because such systems generally show pronounced long-term variability, with X-ray flux changes of up to two order of magnitude, associated with eccentric orbits and changes in the Be-star decretion disk rather than maintaining a persistently high luminosity \citep{2011Ap&SS.332....1R}. 
Moreover, both the SFXT and Be-HMXB scenarios are subject to the difficulty of no seeing a luminous OB donor (Section~\ref{sec:optical}), which is required for sustaining a strong wind and in turn the high mass-accretion rate implied by the persistent high luminosity of M81F4.

This difficulty can be relaxed in a disk-fed mass-transfer geometry. Persistent disk formation is more readily achieved when the donor nearly fills or fills its Roche lobe, allowing Roche-lobe overflow \citep{2002ApJ...565.1107P}, in principle providing a more natural way to sustain the high persistent accretion rate of M81F4 than the accretion of a gravitationally focused wind (see further discussions in Section~\ref{subsec:NS}).

LMC X-4 and SMC X-1 provide an instructive comparison because both are persistently luminous, disk-fed NS-HMXBs that produce short X-ray flares. LMC X-4 shows complex and often multi-peaked flares with a mean FWHM duration of $\sim68$ s. The strongest events can reach $\sim2\times10^{39}\ {\rm erg\ s^{-1}}$ and sometimes occur in clustering separated by $\sim2000$ s. Its spectrum generally softens as the intensity increases \citep{2003ApJ...586.1280M}. SMC X-1 produces shorter and more frequent flares, with a mean FWHM duration of $\sim18$ s and typical separations of $\sim100$ s. These flares are approximately symmetric, reach about $2.5$ times the non-flare intensity, and show little spectral evolution \citep{1997A&A...321L..25L,2003ApJ...582L..91M}. 
For LMC X-4, episodic enhancement of accretion  may be triggered by Rayleigh-Taylor instability at the disk-magnetosphere \citep{2003ApJ...586.1280M}. In SMC X-1, accretion enhancements may arise from a viscous instability near the inner disk edge \citep{2003ApJ...582L..91M}. 
Nevertheless, M81F4 differs from both systems in several important aspects. Its flares last a few hundred seconds and have a fast rise followed by a slower decay. Their averaged recurrence time is $\sim0.5$ d, much longer than the characteristic flare separations in LMC X-4 and SMC X-1. M81F4 also shows no significant hardness evolution during the flares. On longer timescales, LMC X-4 and SMC X-1 exhibit strong superorbital modulation associated with warped, precessing accretion disks \citep{2004ApJ...600..351N,2003MNRAS.339..447C}. No comparable recurrent low states are established for M81F4, which has remained luminous for more than two decades, showing only a moderate long-term variability. LMC X-4 and SMC X-1 contain optically identified massive evolved donors, classified as an O8 III star and a B0 I supergiant, respectively, and both undergo Roche-lobe overflow \citep{2007A&A...473..523V}, while no comparably luminous counterpart is present for M81F4.

\subsection{An Accreting Magnetized Neutron Star?}
\label{subsec:NS}
Based on the observed properties of M81F4, and inspired by the above comparison with known X-ray binaries exhibiting bursting activities, we propose a scenario in which the flares are produced by brief episodes of enhanced accretion onto a magnetized NS, preferentially supplied by a Roche-lobe overflow.
In such a system, matter transferred from the donor forms an accretion disk which is truncated at the magnetospheric radius $r_{\rm m}=\xi(\frac{B^4R_{\rm NS}^{12}}{2GM_{\rm NS}\dot M^2})^{1/7}$, where the magnetic pressure begins to dominate the accretion flow. Here $B$ is the NS surface dipole field strength at the magnetic equator, $R_{\rm NS}$ is the NS radius, $M_{\rm NS}$ is the NS mass, $\dot M$ is the mass inflow rate at the inner disk, and $\xi$ is an order unity dimensionless factor describing the interaction between the disk and magnetosphere \citep{2017MNRAS.470.2799C}. Inside this radius, the accreting material is channeled along the magnetic field lines towards the magnetic poles, forming accretion funnels or columns from which the bulk of the hard X-rays are produced \citep{2025MNRAS.536..340K}. 

Consider the case in which the inner disk radius, set by the magnetosphere, lies close to the co-rotation radius $r_{\rm co}=(\frac{GM_{\rm NS}P_{\rm spin}^2}{4\pi^2})^{1/3}$,  where the Keplerian angular velocity equals the NS spin angular velocity defined by the spin period $P_{\rm spin}$. 
Under suitable condition, this configuration can support the RII accretion instability described by \citet{2012MNRAS.420..416D}. In the RII regime, the magnetic torque exerted on the disk varies sharply across a narrow transition near co-rotation. This strong torque response couples small changes in the inner disk radius to variations in the accretion rate, providing the feedback that drives the instability. 
In this case, the accretion does not need to switch abruptly between completely ``off'' and ``on'' states. 
Instead, a magnetically gated, partially leaky accretion flow allows the inflowing material to continue accreting onto the NS, thereby sustaining the persistent high luminosity. Meanwhile, a small fraction of matter can temporarily accumulate in the innermost disk near co-rotation when the local mass supply exceeds the mass accretion rate onto the NS. This allows a small reservoir of gas to build up near the magnetospheric boundary while accretion continues. 
Eventually a further increase in the local mass inflow rate pushes the inner disk edge inward, overcoming the magnetospheric barrier. This allows the locally stored gas to accrete suddenly onto the NS, producing a flare \citep{2010MNRAS.406.1208D,2012MNRAS.420..416D}. 
As the local reservoir is depleted, the enhanced accretion subsides and the inner disk edge moves outward again, allowing the luminosity to decline while accretion onto the NS continues.

In this scenario, the flare energy is supplied by the prompt release of gravitational potential energy during episodes of enhanced accretion onto the NS, in contrast to Type I bursts that are powered by nuclear fusion on the NS surface. 
Assuming $M_{\rm NS} = 1.4M_\odot$ and $R_{\rm NS} =$ 10 km, for the observed persistent luminosity $L_{\rm pers}\simeq6\times10^{38}~{\rm erg~s^{-1}}$, an average flare peak luminosity $L_{\rm peak}\simeq3\times10^{39}~{\rm erg~s^{-1}}$, and a characteristic duration $\Delta t\simeq200$ sec, the excess flare energy corresponds to a stored mass of only $\Delta M_{\rm exc} = (L_{\rm peak}-L_{\rm pers})\Delta tR_{\rm NS}/(GM_{\rm NS} \eta_{\rm rad}) \sim1.3\times10^{-12}\eta_{\rm rad}^{-1}~M_\odot$, where $\eta_{\rm rad}$ parameterizes the fraction of the accretion power radiated in the observed X-ray band. 
For an average recurrence time of $t_{\rm rec}\sim$ 0.5 day, the implied average storage rate is therefore $\dot M_{\rm store} = \Delta M_{\rm exc}/t_{\rm rec} \sim10^{-9}\eta_{\rm rad}^{-1}~M_\odot~{\rm yr^{-1}}$ (for $t_{\rm rec}\sim$ 6 ks between the two nearest flares, $\dot M_{\rm store} \sim10^{-8}\eta_{\rm rad}^{-1}~M_\odot~{\rm yr^{-1}}$). For comparison, the persistent luminosity corresponds to an accretion rate of $\dot M_{\rm pers}=L_{\rm pers}R_{\rm NS}/(GM_{\rm NS}\eta_{\rm rad})\simeq 5\times10^{-8}\eta_{\rm rad}^{-1}M_{\odot}\rm~yr^{-1}$. Thus, the material that must be temporarily stored near the magnetospheric boundary is only $\sim 2\%-20\%$ of the persistent accretion flow, supporting the notion that the magnetosphere does not need to halt the entire inflow, but only needs to modulate a small fraction of the steady, high accretion rate. Conversely, a substantial reduction of the mass supply rate to the inner disk should reduce or even halt the mass storage (in the meantime $r_m$ becomes larger) and the subsequent flare occurrence.
The non-detection of flares in the 2022 {\it Chandra} observation and 2024 {\it XMM-Newton} observation (Figure~\ref{fig:long}), coincident with a substantial decrease of the persistent luminosity, might be understood in this regard. 

The mean persistent luminosity of $\sim6\times10^{38}~{\rm erg~s^{-1}}$ and the flare peak luminosity of $\sim3\times10^{39}~{\rm erg~s^{-1}}$ correspond to approximately 3 and 15 times the Eddington limit of a canonical $1.4~M_\odot$ NS, respectively. 
Although the inferred accretion rate is super-Eddington in M81F4, the disk may remain geometrically thin if the NS magnetosphere truncates it outside the region where substantial thickening is expected. In this configuration, much of the accretion power can be released in the accretion column inside the truncation radius, allowing the total luminosity to exceed the Eddington limit without requiring a geometrically thick disk \citep{2017MNRAS.470.2799C}. Under these conditions, the thin-disk approximation underlying the D'Angelo--Spruit model could remain applicable, and its RII regime remains a possible explanation for the flares.

More quantitatively,
the disk is truncated at the magnetospheric radius, $r_{\rm m}\simeq6\times10^7~\xi B_{12}^{4/7}(\eta_{\rm rad}/0.1)^{2/7}~{\rm cm}$, where $B_{12}=B/(10^{12}~{\rm G})$. For the illustrative values $B_{12}=1$, $\eta_{\rm rad}=0.1$, and $\xi=1$, this gives $r_{\rm m}\sim6\times10^7~{\rm cm}$. 
For an electron-scattering opacity of $\kappa=0.34~{\rm cm^2~g^{-1}}$, these parameters correspond to $\dot m_{\rm per}\equiv\dot M_{\rm per}c^2/L_{\rm Edd}\simeq 140$, where $L_{\rm Edd}=4\pi GM_{\rm NS}c/\kappa$. The corresponding spherization radius is $r_{\rm sph}\simeq\frac{3}{2}\dot m_{\rm per}\frac{GM_{\rm NS}}{c^2}\simeq4\times 10^7 \rm~cm$, which marks the characteristic radius inside which radiation pressure can substantially thicken the disk \citep{2017MNRAS.470.2799C}. Thus, given a sufficiently strong magnetic field ($B_{12} \sim 1$), the estimated $r_{\rm m}$ would be larger than $r_{\rm sph}$, allowing the disk to be truncated before reaching the region where strong geometric thickening is expected. The fiducial calculation of \citet{2017MNRAS.470.2799C} also yields a maximum relative disk thickness of order of 0.25 for characteristic magnetic moments of NS, $\mu\equiv BR_{\rm NS}^{3}
\sim1\times 10^{30}~{\rm G\,cm^{3}}$ and $\dot m\sim100$. This demonstrates that a relatively thin disk is possible at an accretion rate comparable to M81F4. 

If the flare decay reflects the draining of material accumulated near the inner disk, its duration can be compared with the local viscous timescale $t_{\rm visc}\simeq\alpha^{-1}(H/r_{\rm m})^{-2}\Omega_{\rm K}^{-1}$, where $\Omega_{\rm K}=(\frac{GM_{\rm NS}}{r_{\rm m}^3})^{1/2}$. Adopting $\alpha=0.1$ and $H/r_{\rm m}=0.1$ as illustrative values gives $t_{\rm visc}\simeq34\xi^{3/2}B_{12}^{6/7}(\eta_{\rm rad}/0.1)^{3/7}(\alpha/0.1)^{-1}(\frac{H/r_{\rm m}}{0.1})^{-2}~\rm s$, which is comparable in order of magnitude to the observed flare decay time of $\sim200~{\rm s}$.

Equating $r_{\rm m}$ with $r_{\rm co}\simeq1.7\times10^{8}~(P_{\rm spin}/1\rm~s)^{2/3}\rm~cm$ gives $P_{\rm spin}\sim0.2~\xi ^{3/2}B_{12}^{6/7}(\eta_{\rm rad}/0.1)^{3/7}\rm~s$. Taking typical value of $\eta_{\rm rad}=0.1$ and $\xi=1$, this corresponds to $P_{\rm spin}\sim 0.2-1.4~s$ for $B\sim10^{12}-10^{13}\rm~G$, although neither the magnetic field nor the spin period is currently constrained. 
We note that no significant periodic or quasi-periodic variability is detected over $1-20000$ sec, following the procedure detailed in \cite{2026MNRAS.546f2237Z}. 
In particular, the sensitivity to periodic signal detection at $\lesssim$1 sec is severely limited by photon statistics.  
Thus, the non-detection does not rule out the proposed spin-period range.

We emphasize that the apparently super-Eddington luminosities do not rule out a magnetized-NS accretor, as ultra-luminous X-ray pulsars (ULXPs) can reach persistent luminosities of $10^{39}-10^{40}~{\rm erg~s^{-1}}$, and occasionally even higher, through magnetic channeling, anisotropic radiation, and enhanced photon escape from the accretion column \citep{2014Natur.514..202B,2017Sci...355..817I,2002ApJ...568L..97B}. However, no ULXP is known to exhibit M81F4-like flares. Moreover, known ULXPs commonly exhibit substantial long-term variations, sometimes exceeding several orders of magnitude, whereas M81F4 shows modest long-term variability \citep{2020MNRAS.491.1260S}.

The X-ray spectra further support a magnetized NS origin. Both the flare and non-flare states are dominated by hard power-law continua, with no additional thermal component required during the flares, suggesting a temporary increase in the accretion rate without a substantial change in the accretion geometry. The hard X-ray emission is consistent with magnetically channeled accretion onto an NS, in which seed photons from bremsstrahlung, cyclotron emission, and thermal radiation are up-scattered through bulk or thermal Comptonization in the accretion column \citep{2007ApJ...654..435B}, similar to the case of ULXPs. 
The modest spectral softening during the flares may result from the increased seed-photon flux and more efficient Compton cooling at higher accretion rates \citep{2020ApJ...895..114H}. The 6.7 keV emission feature further indicates the presence of highly ionized material in the persistent accretion flow, plausibly located in the inner-disk atmosphere or near the disk–magnetosphere interface and irradiated by the central X-ray source.
The weak soft/thermal component in the non-flare spectrum may arise from reprocessing of the hard X-rays by optically thick material in the inner disk or near the disk-magnetosphere interface, as commonly invoked in luminous and ultraluminous X-ray pulsars \citep{2025MNRAS.536..340K}.

In the above scenario, a massive main-sequence or supergiant donor is not necessary to sustain the high accretion rate. 
Instead, the donor can be an intermediate-mass star with a few $M_{\odot}$, which supplies material through Roche lobe overflow and forms with the compact accretor a so-called intermediate-mass X-ray binary (IMXB). 
Such systems can undergo rapid mass transfer on thermal timescale of the donor, thereby sustaining a high mass transfer rate while naturally forming an accretion disk \citep{2002ApJ...565.1107P,2007A&A...476.1283X}. 
Moreover, IMXBs can undergo a relatively short-lived ($\sim$1 Myr) phase of thermal-timescale mass transfer, which may help explain the scarcity of M81F4-like systems \citep{2002ApJ...565.1107P}. 
Interestingly, among the candidate optical counterparts, several sources are consistent with being $4-5\rm~M_\odot$ stars in the subgiant or red giant phase (Section~\ref{sec:optical}; Figure~\ref{fig:hst}), which satisfies the IMXB condition. 
This scenario is also consistent with the lack of a massive optical counterpart and may help explain the absence of a detectable neutral Fe K$\alpha$ line. In contrast to a wind-fed system with a massive donor, Roche lobe overflow does not require a dense, relatively cool stellar wind, thereby reducing the amount of low-ionization material available to produce the 6.4 keV fluorescence line. 

On the other hand, a low-mass or white-dwarf donor is disfavored by the X-ray spectral properties of M81F4. The non-flare spectrum, with $\Gamma\simeq 1.1$, is substantially harder than that of typical luminous LMXBs (with $\Gamma\simeq1.6$; \citealp{2003ApJ...587..356I}). 
NS-LMXBs exhibiting a luminous hard continuum are often found as the so-called {\it Z sources}, which typically contain thermal or low-temperature Comptonized emission and show spectral evolution with luminosity \citep{2012ApJ...756...34L}. 
This contrasts with the lack of detectable spectral evolution in M81F4. 
A white-dwarf donor would imply an ultra-compact X-ray binary (UCXB). However, the mean persistent luminosity of M81F4, $6\times10^{38}~\rm erg~s^{-1}$, significantly exceeds that of the brightest known UCXB candidates ($3\times10^{38}~\rm erg~s^{-1}$) \citep{2024MNRAS.530.2096Z,2026Innov...701181M}.

\subsection{Accreting Black Hole Systems}
Lastly, we consider the alternative possibility that M81F4 involves an accreting stellar-mass black hole. 
For a $10-20~M_\odot$ BH, the persistent luminosity would remain sub-Eddington, while the flare peaks would approach or modestly exceed the Eddington limit. Galactic BH-XRBs such as GRS~1915+105 and V4641~Sgr show rapid luminous variability driven by unstable accretion flows \citep{1999ApJ...517..935Y,2000ApJ...528L..93W}, but neither of them closely resembles M81F4. 
In particular, the large-amplitude outbursts and pronounced spectral-state evolution commonly seen in BH transients \citep{2010MNRAS.403...61D} are highly dissimilar to M81F4. 
A stellar-mass BH interpretation would therefore require an atypical persistent system with recurrent variability in the inner accretion flow. An accreting intermediate-mass black hole (IMBH) also remains a plausible alternative, although the origin of the recurrent flares would remain unclear in this scenario.

\section{Summary}

We have characterized M81F4, a persistently luminous X-ray source in M81 with recurrent short flares. Our main findings are as follows:

\begin{enumerate}
\item M81F4 has remained luminous over two decades, with a typical 0.5-8.0 keV luminosity of $10^{38}-10^{39}\rm~erg~s^{-1}$.

\item We identify 21 flares lasting a few hundred seconds and reaching peak luminosities of a few $10^{39}\rm~erg~s^{-1}$. Their mean recurrence time is $\sim0.5$ day, with a minimum observed separation of $\sim6\rm~ks$.

\item The stacked flare profile shows fast-rise and slow-decay  without significant hardness evolution. Both flare and non-flare spectra exhibit similarly hard continua, with no additional thermal component required during the flares.
\end{enumerate}

M81F4 therefore occupies a rare observational regime in which recurrent minute-scale luminous flares are superimposed on an already high and persistent accretion state. 
The flare energetics and lack of spectral evolution disfavor a classical thermonuclear origin and instead suggest brief enhancements in a sustained accretion. We favor an accreting, magnetized NS, possibly fed through Roche-lobe overflow from an intermediate-mass donor. In this picture, a small fraction of the inflowing material is temporarily stored near the disk-magnetosphere boundary and released episodically to produce flares, while the bulk flow continues to accrete steadily. 

If confirmed, this interpretation would make M81F4 a rare probe of disk-magnetosphere interaction at persistently high and apparently super-Eddington luminosities. 
This discovery highlights the potential of extragalactic time-domain studies to reveal phenomena absent or under-represented in Galactic samples. Searches for analogs can constrain their occurrence rate and environmental dependence, while providing observational benchmarks for theoretical models. 

\begin{acknowledgments}
This work is supported by the National Natural Science Foundation of China (grant No. 12225302, No. 12121003, No. 12473041) and the Fundamental Research Funds for the Central Universities (grant KG202502). 
The authors wish to thank Drs. Nathalie Degenaar, Jialai Kang, Fangzheng Shi, and Zhao Su for helpful discussions. The authors thank the anonymous referee for helpful comments that improve our work.

The authors used ChatGPT (OpenAI) to refine and correct the manuscript to improve the clarity of the language. All AI-assisted text was reviewed and verified by the authors for scientific accuracy. Responsibility for the final manuscript and its scientific content lies entirely with the authors.

\end{acknowledgments}
\bibliography{sample701}{}
\bibliographystyle{aasjournalv7}

\newpage
\appendix
\renewcommand*{\theHequation}{appendix.\thesection.\arabic{equation}}
\makeatletter
\@addtoreset{table}{section}
\@addtoreset{figure}{section}
\makeatother

\renewcommand{\thetable}{\thesection\arabic{table}}
\renewcommand{\thefigure}{\thesection\arabic{figure}}

\section{Individual flare profiles}
\label{app:individual}
\begin{figure*}[h]
    \centering

    \includegraphics[width=0.32\textwidth]{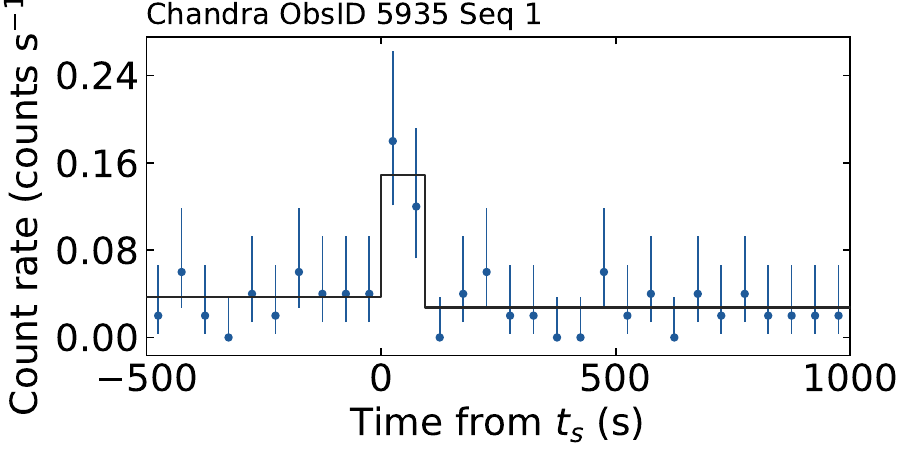}
    \includegraphics[width=0.32\textwidth]{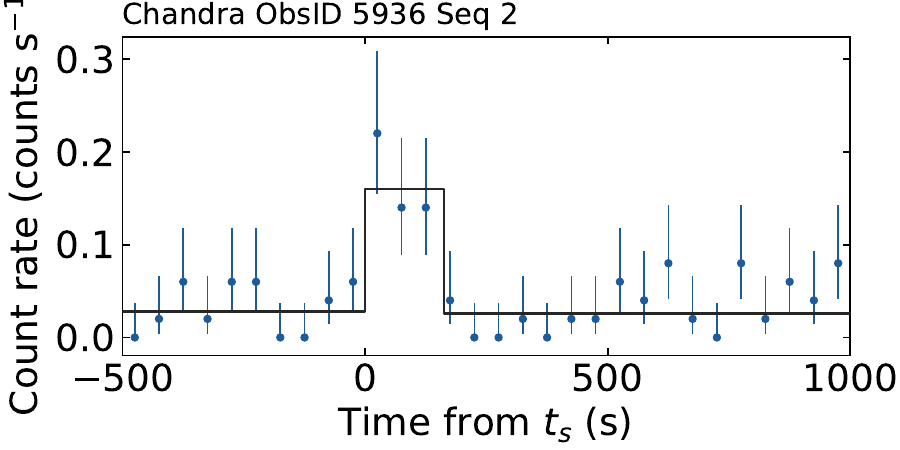}
    \includegraphics[width=0.32\textwidth]{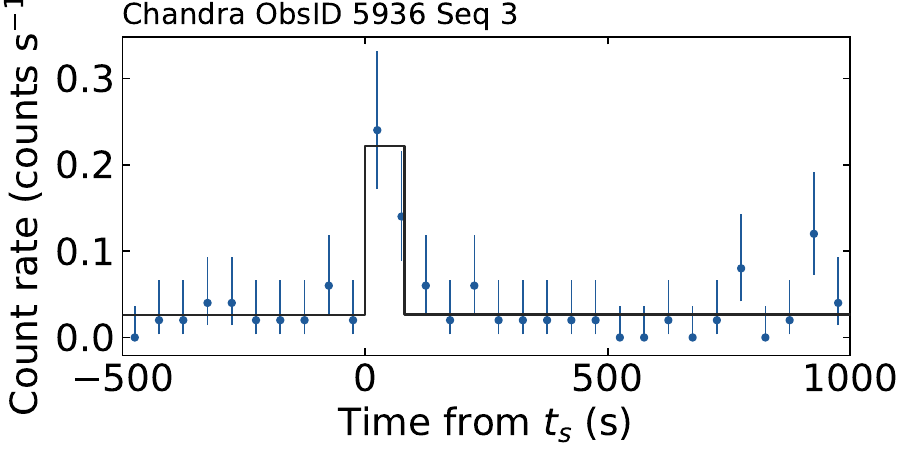}\\

    \includegraphics[width=0.32\textwidth]{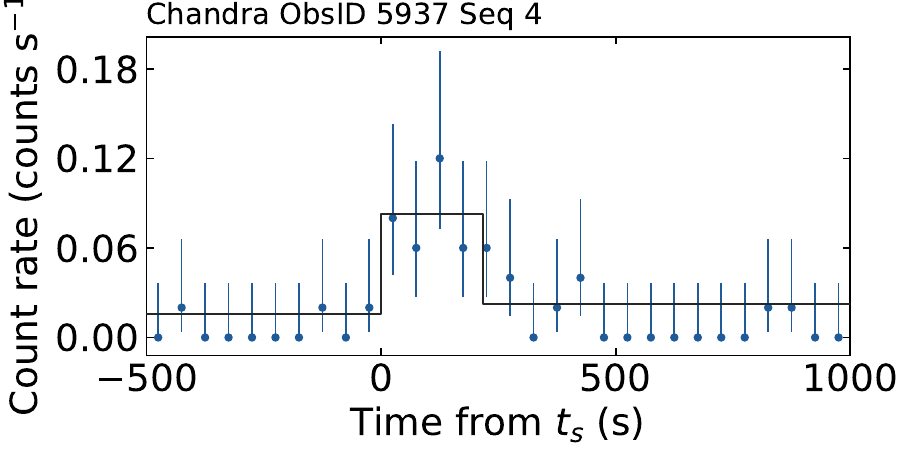}
    \includegraphics[width=0.32\textwidth]{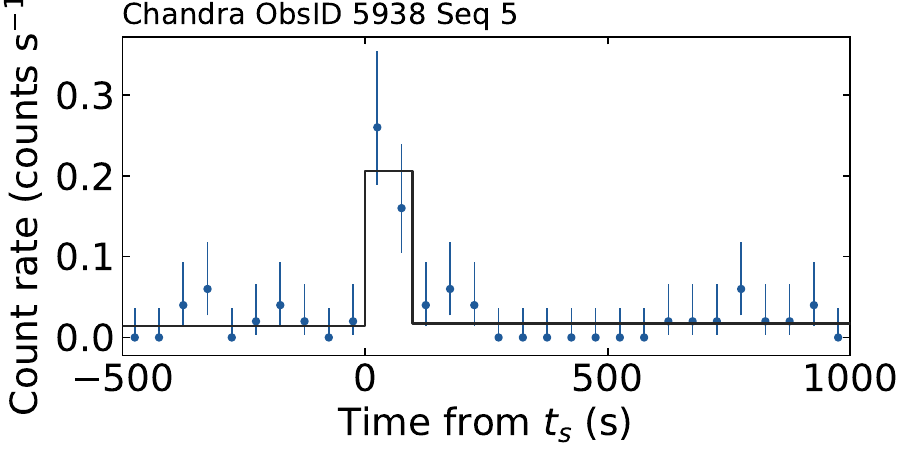}
    \includegraphics[width=0.32\textwidth]{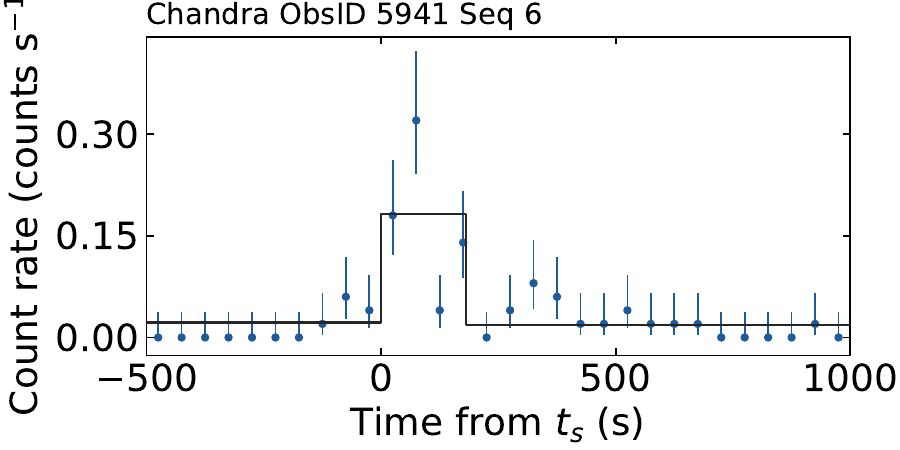}\\

    \includegraphics[width=0.32\textwidth]{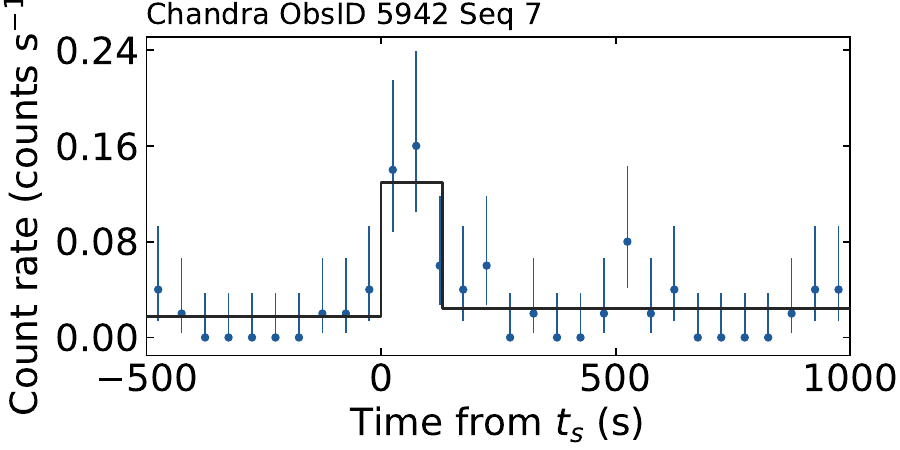}
    \includegraphics[width=0.32\textwidth]{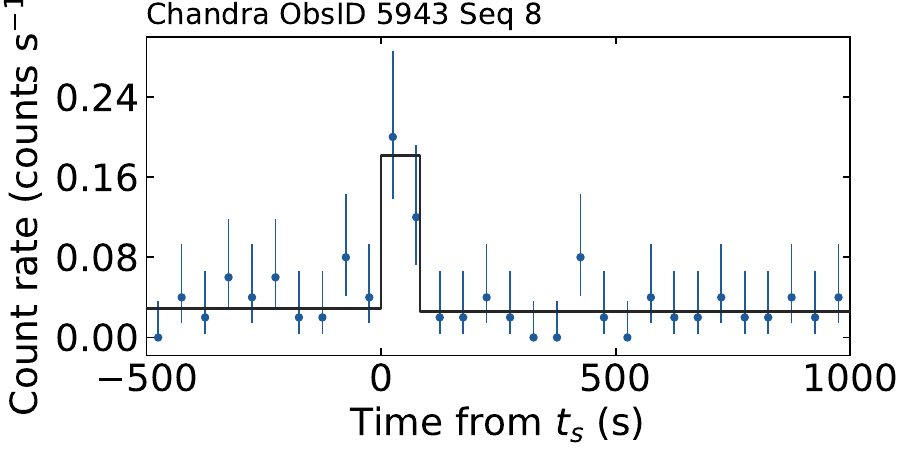}
    \includegraphics[width=0.32\textwidth]{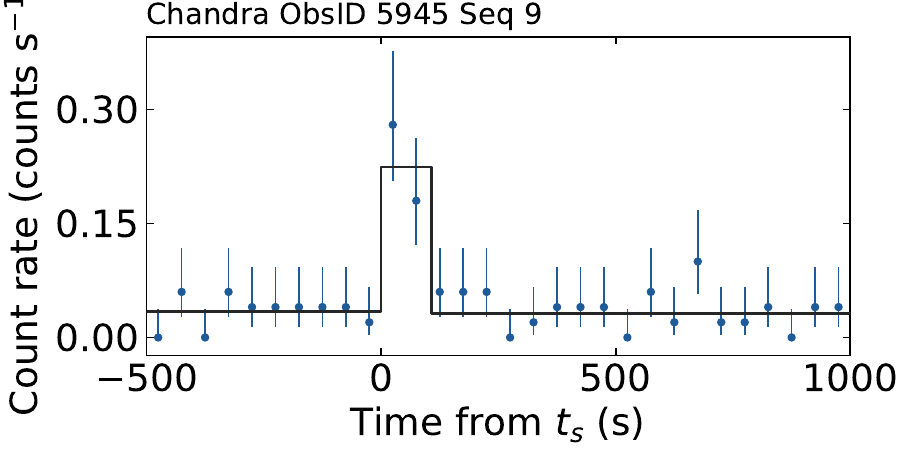}\\

    \includegraphics[width=0.32\textwidth]{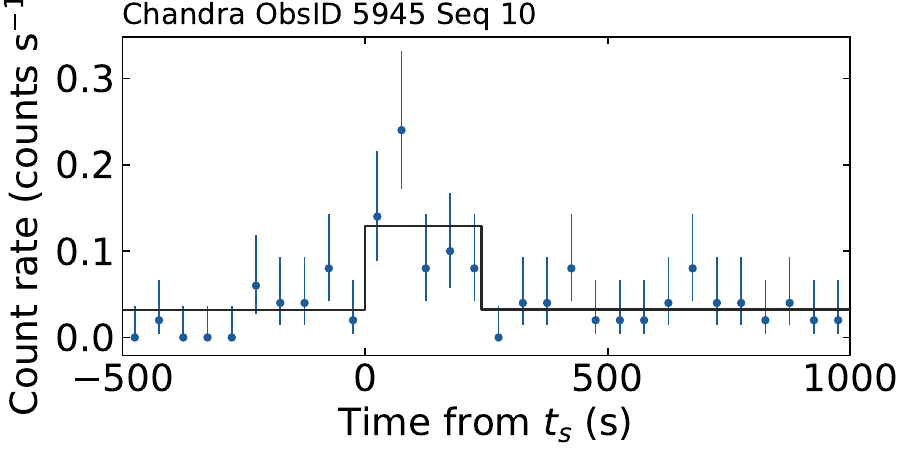}
    \includegraphics[width=0.32\textwidth]{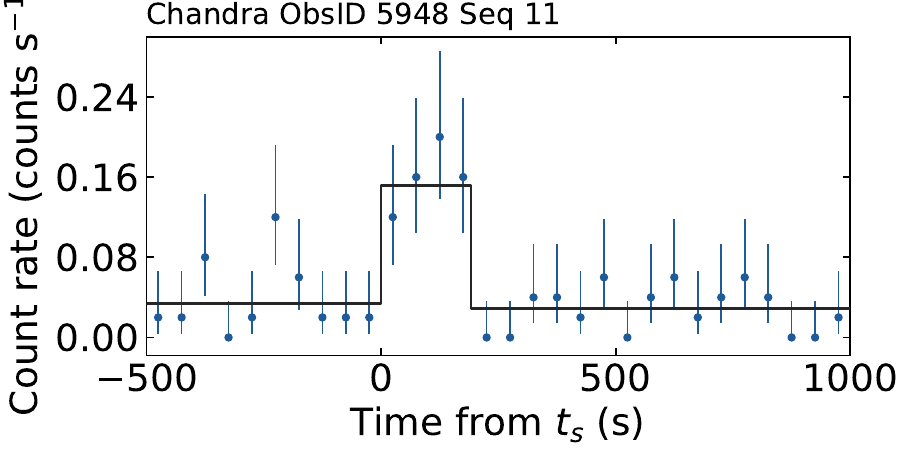}
    \includegraphics[width=0.32\textwidth]{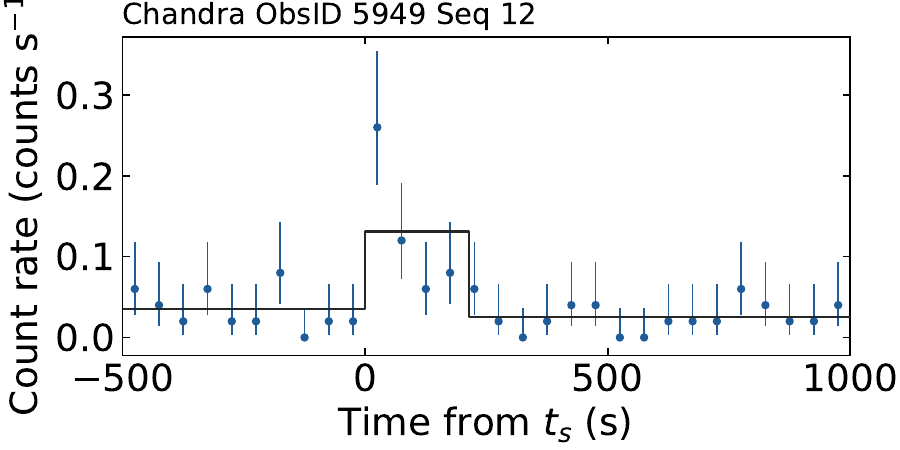}\\

    \includegraphics[width=0.32\textwidth]{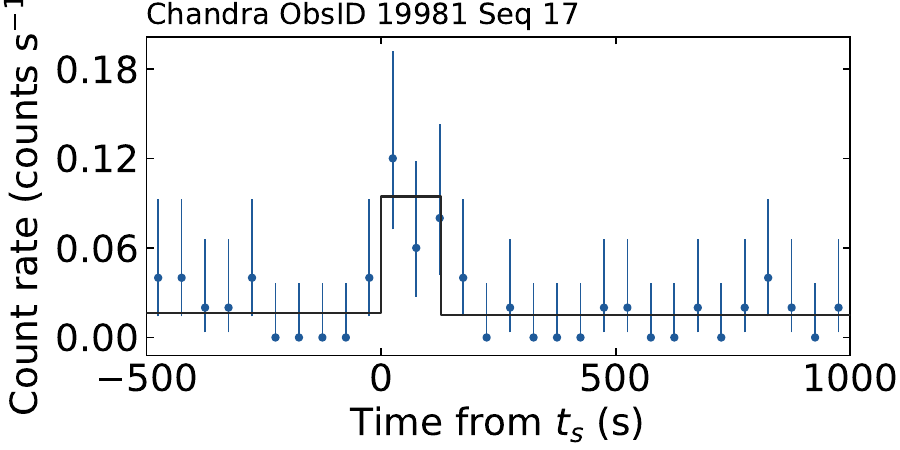}
    \includegraphics[width=0.32\textwidth]{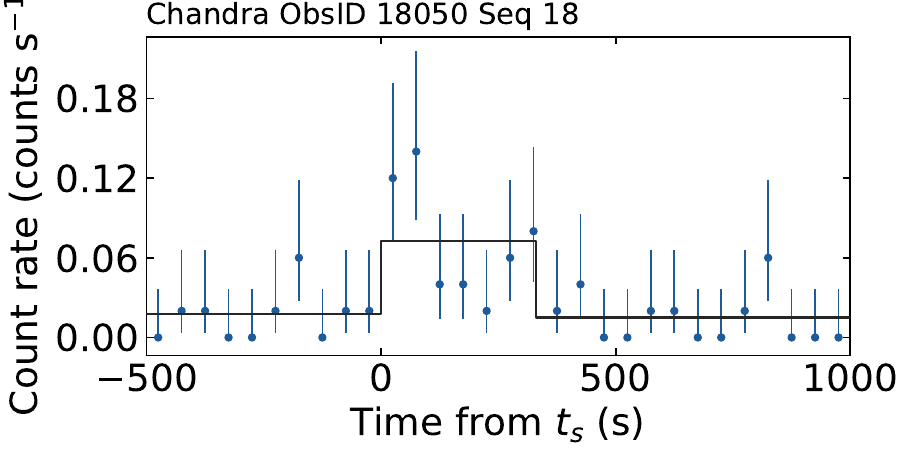}
    \includegraphics[width=0.32\textwidth]{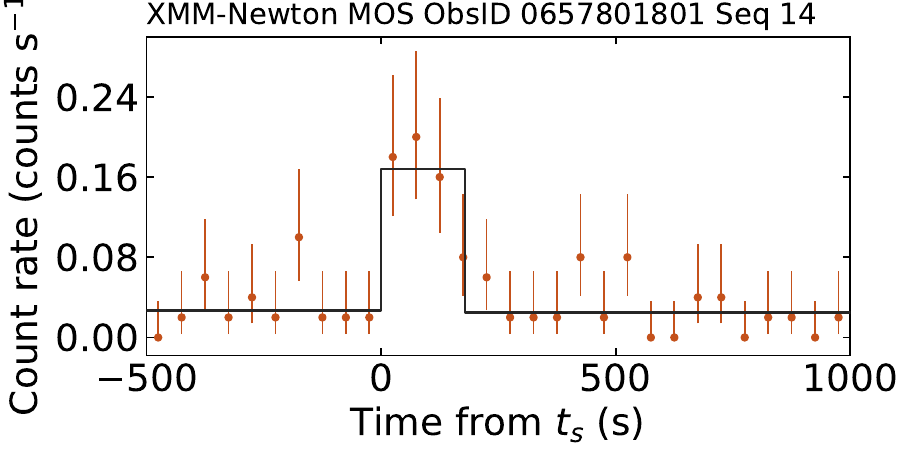}\\

    \includegraphics[width=0.32\textwidth]{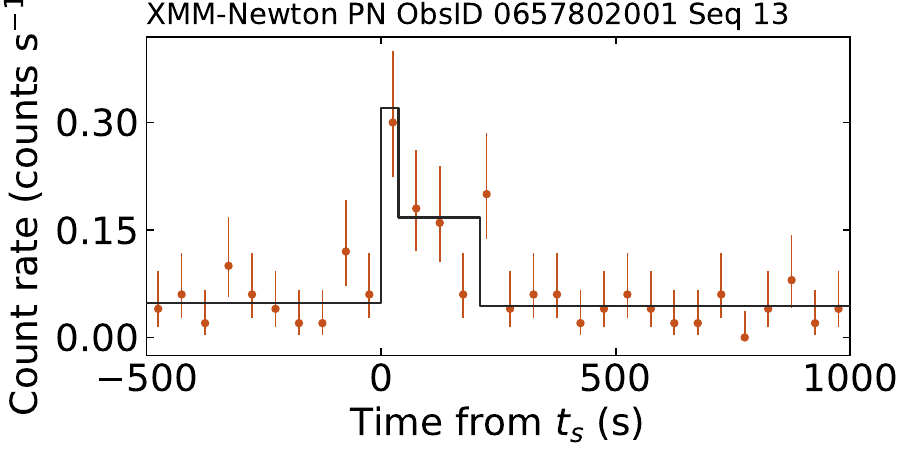}
    \includegraphics[width=0.32\textwidth]{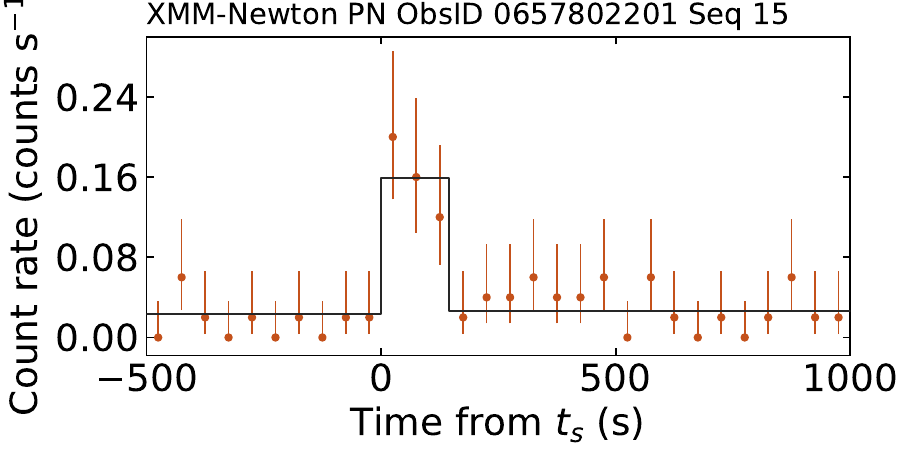}
    \includegraphics[width=0.32\textwidth]{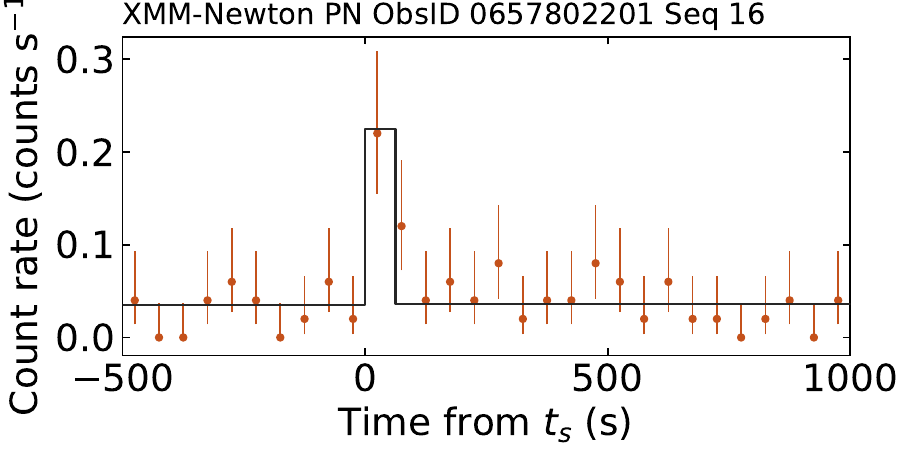}\\

    \includegraphics[width=0.32\textwidth]{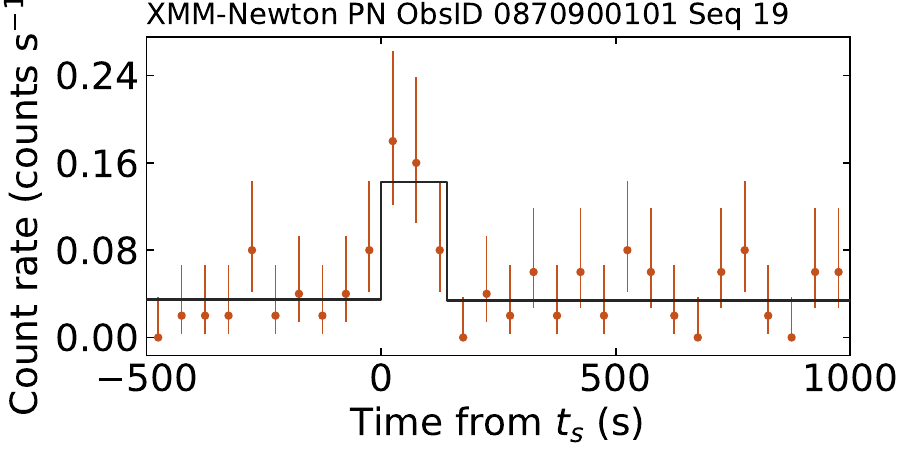}
    \includegraphics[width=0.32\textwidth]{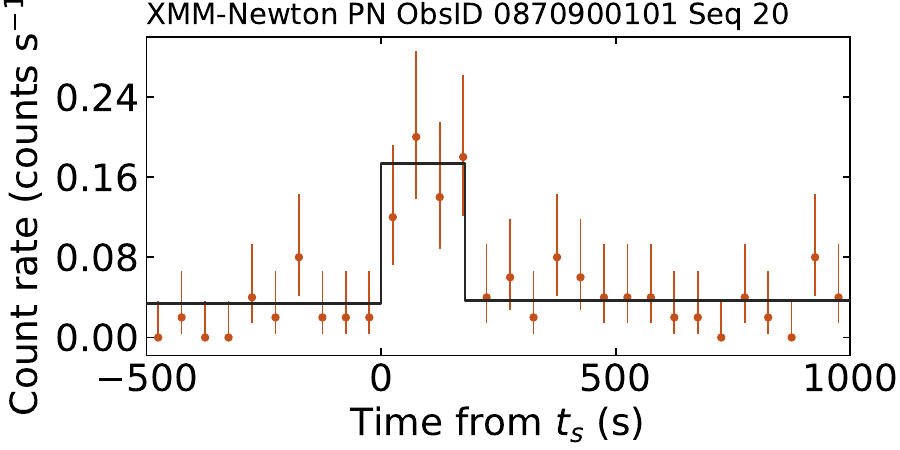}
    \includegraphics[width=0.32\textwidth]{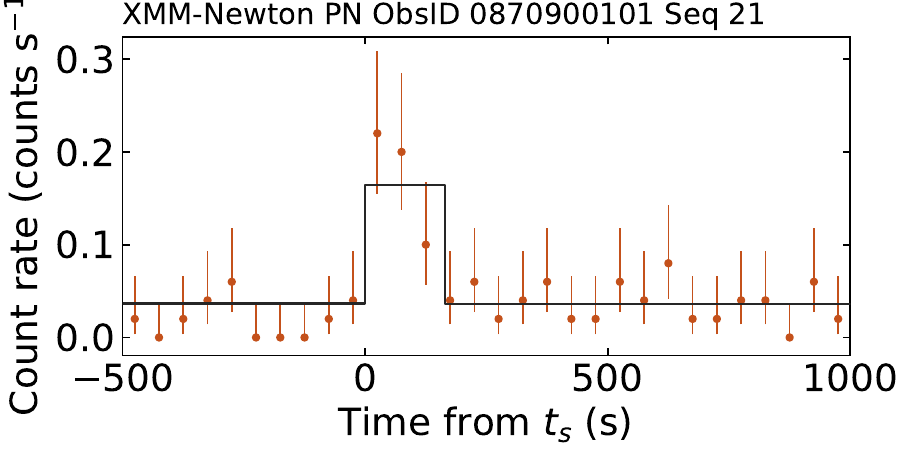}
\caption{0.5--8.0~keV X-ray light curves of the individual flares detected in M81F4. The light curves are shown with uniform 50~s bins over $-500$ to $+1000$~s relative to the Bayesian-block (solid lines) start time of each flare. Error bars represent central 68.3\% Poisson confidence intervals. Most flares show a relatively rapid rise followed by a more gradual decline. 
}
\label{fig:individual_flare_lightcurves}
\end{figure*}

We present the individual flare light curves in Figure~\ref{fig:individual_flare_lightcurves}. The light curves are constructed using uniform 50~s bins over $-500$ to $+1000$~s relative to the Bayesian-block start time of each flare. The flare intervals identified by the Bayesian Blocks algorithm are shaded, with their start and end times marked. Error bars represent central 68.3\% Poisson confidence intervals.

Despite the limited photon statistics, some of the individual flares appear to exhibit a relatively rapid rise followed by a more gradual decay, consistent with the FRED trend seen in the stacked profiles. Dividing each flare block into two equal-duration halves, we find that 15 of the 21 flares ($71.4\%$) have a higher count rate in the first half, supporting this overall trend.

\section{Detection Completeness of the flare search}
\label{app:completeness}
We performed photon level injection-recovery simulation to quantify the completeness of our flare search as a function of flare amplitude, persistent source emission, and background level. 

Each simulated observation spans 10 ks and contains a single fast-rise exponential-decay (FRED) flare with its peak at $t_{\rm pk}=5000$~s. The source count rate is described by
\begin{equation}
R(t)=C+A~g(t),
\end{equation}
where
\begin{equation}
g(t)=
\begin{cases}
\displaystyle
\lambda\exp\left[
-\frac{\tau_1}{t-t_s}
-\frac{t-t_s}{\tau_2}
\right], & t>t_s,\\[6pt]
0, & t\leq t_s.
\end{cases}
\end{equation}
Here, $C$ is the persistent source count rate and $A$ is the peak flare count rate above the persistent emission; the total source count rate at the peak is therefore $C+A$. We adopt $\tau_1=671$~s and $\tau_2=42$~s from the fitted {\it Chandra} FRED profile, and set $t_s=t_{\rm pk}-\sqrt{\tau_1\tau_2}$. This template has a full width at half maximum of $143$~s and a width of $318$~s at $e^{-3}$ of the net peak amplitude.

We consider $A=0.02$, $0.05$, $0.10$, $0.20$, $0.30$, and $0.40~{\rm count~s^{-1}}$, with $C/A=0.2$ or $0.5$. The raw count rate in the independent background aperture, $B$, is set to $B/C=0.05$, giving 12  combinations of parameters. The background aperture has 12 times the source aperture area, so $B_{\rm src}=B/12$. For each combination, we generate 1000 simulated observations containing an injected flare and 1000 independent no-flare controls. We first generate persistent source photons and source-aperture background photons as independent homogeneous Poisson processes with rates $C$ and $B/12$, respectively. We then add flare photons generated from an inhomogeneous Poisson process with rate $A\,g(t)$. These three independent event lists are merged and sorted to form the source-aperture photon arrival times. The background-aperture event list is generated independently as a homogeneous Poisson process with rate $B$. 
The source-aperture event list combines all three components, and the background-aperture event list is generated independently at rate $B$. Continuous photon arrival times and fixed random seeds are used throughout.

The event lists are searched using our implementation of the two-step Bayesian Blocks framework \citep{2026MNRAS.546f2237Z}, with background-dependent time correction before the second segmentation. Blocks with rates above that of the longest block and binomial probabilities below $2.7\times10^{-3}$ are selected, and adjacent qualifying blocks are merged into candidate intervals. An injected flare is recovered when a candidate interval contains its true peak time. We define the completeness as $\epsilon=N_{\rm recovered}/1000$. In the no-flare controls, any candidate interval constitutes a false detection, giving an observation-level false detection rate $f_{\rm false}=N_{\rm false}/1000$. 
\begin{figure*}
\centering
\includegraphics[width=\textwidth]{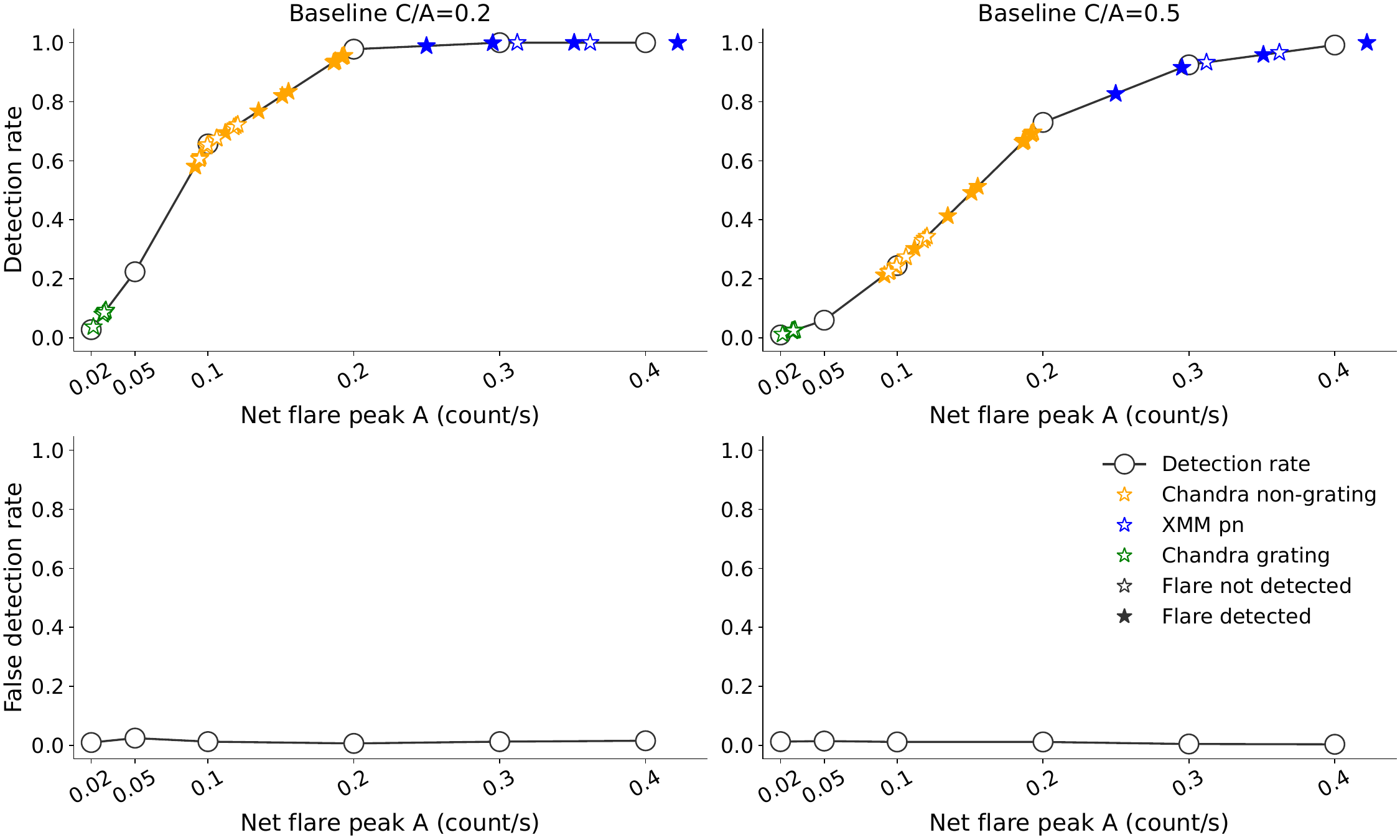}
\caption{Detection completeness (upper panels) and false detection rate (lower panels) of the two-step Bayesian Blocks search versus net flare peak count rate. Columns show $C/A=0.2$ (left) and $0.5$ (right). Stars mark observation-specific count rates for the adopted common net peak luminosity. Blue, orange, and green stars denote {\it Chandra} non-grating, {\it XMM--Newton} pn, and {\it Chandra} grating observations, respectively. Filled and open stars identify ObsIDs with and without detected flares. }
\label{fig:completeness}
\end{figure*}
Figure~\ref{fig:completeness} shows the resulting detection completeness and false detection rates. We adopt $C/A=0.2$ as the fiducial configuration, which most closely resemble the observed source and background levels. Here $B/C=0.05$ refers to the unscaled background-aperture rate, corresponding to $B_{\rm src}/C=0.05/12\simeq0.004$. For this configuration, the completeness is $2.8\%$, $22.4\%$, $65.7\%$, $97.8\%$, $100.0\%$, and $100.0\%$ at $A=0.02$, $0.05$, $0.10$, $0.20$, $0.30$, and $0.40~{\rm count\,s^{-1}}$, respectively. The corresponding false detection rates are $1.0\%$, $2.5\%$, $1.3\%$, $0.7\%$, $1.3\%$, and $1.6\%$. The entries of $100\%$ indicate recovery of all 1000 injected flares. Figure~\ref{fig:completeness} shows the resulting detection completeness and false detection rates. A stronger persistent component reduces the recovery fraction: at  $C/A=0.5$, the completeness is $70.4\%$ at $A=0.2~{\rm count~s^{-1}}$ and $94.2\%$ at $A=0.3~{\rm count~s^{-1}}$. We also tested $B/C=0.1$ and $0.2$. These background levels have only a small effect on the detection rate within the tested range, consistent with the low background contribution in the source aperture.

To compare the simulations with the observations, we assume that all flares have the same net peak luminosity, $L_{\rm pk}=3\times10^{39}~{\rm erg\,s^{-1}}$, in the $0.5-8$~keV band. The expected net peak count rate for observation $i$ is
\begin{equation}
A_i=\frac{L_{\rm pk}}{4\pi d^2}\frac{\mathcal{A}_{{\rm eff},i}}{k_{\rm ph}},
\end{equation}
where $d=3.6$~Mpc, $\mathcal{A}_{{\rm eff},i}$ is the effective area for observation $i$, and $k_{\rm ph}\simeq3.52\times10^{-9}~{\rm erg\,photon^{-1}}$ converts photon flux to energy flux in the same band. We then linearly interpolate the simulated completeness at $A_i$ using the fiducial curve.

For the {\it Chandra} non-grating observations, the assumed peak luminosity corresponds to $A_i=0.091-0.193~{\rm count~s^{-1}}$, giving interpolated completeness values of approximately $57.4-95.1\%$. For the {\it Chandra} grating observations, the estimated detection completeness at the same assumed peak luminosity is $\lesssim10\%$, which explains the absence of detected flares in these observations. For the 6 retained {\it XMM-Newton} PN observations, $A_i=0.250-0.422~{\rm count~s^{-1}}$, corresponding to completeness values of approximately $98.6-100.0\%$. The MOS2 data from ObsID~0657801801 give $A_i=0.091~{\rm count~s^{-1}}$ and a completeness of approximately $56.9\%$, but still have a flare detected.
The same flare, however, was not captured in the PN data due to dead time.

We use these completeness estimates to approximately correct the number of flares and mean recurrence times. For $n_i$ detected flares  and completeness $\epsilon_i$ in observation $i$, the corrected number of flares is $\widehat{N}_{\rm corr}=\sum_i n_i/\epsilon_i$. The 14 flares detected in the 24 {\it Chandra} observations with response conversions correspond to approximately 18 flares at full completeness, whereas the 7 flares in the six {\it XMM-Newton} PN observations require a negligible correction ($\widehat{N}_{\rm corr}\simeq7.76$). The corrected mean recurrence time in each epoch is estimated as $\widehat{t}_{\rm rec}=\sum_i T_i/\widehat{N}_{\rm corr}$, where $T_i$ is the exposure of observation $i$ and the sum includes observations with no detected flares. 
For the full sample of 24 {\it Chandra} observations, the mean recurrence time decreases from $0.5$ to $0.4~{\rm day}$ after completeness correction, whereas for all eight {\it XMM-Newton} observations, with a total exposure of $315.6~{\rm ks}$, it remains unchanged at $0.5~{\rm day}$.


\section{HST images and photometry of candidate counterparts}
\label{app:optical}

Archival HST imaging data were used to search for potential optical counterparts of M81F4. The observations were obtained with the ACS/WFC instrument in the F435W, F606W, and F814W filters, with exposure times of 1200 sec, 1200 sec, and 1650 sec, respectively (datasets J9EL07010, J9EL07020, and J90LA3010). 
Point-source photometry was performed on the CTE-corrected \texttt{FLC} exposures using the ACS module of \texttt{DOLPHOT}, with the drizzled \texttt{DRC} images adopted as astrometric reference frames. We selected high-quality stellar sources by requiring ${\rm S/N}>4$, $\texttt{sharpness}^2<0.2$, and $\texttt{crowding}<2.25$. Sources satisfying these criteria are referred to as good star candidates.

We determine the X-ray centroid of M81F4 using the {\it Chandra} observation ObsID 21384, in which  M81F4 has the smallest off-axis angle ($\sim$$15^{\prime\prime}$) and thus the sharpest PSF. 
The centroid is measured to be RA=09:55:24.22, Dec=+69:09:57.95 (J2000) with a statistical uncertainty of $0.07^{\prime\prime}$ at 95\% confidence level. Thus the X-ray positional uncertainty is dominated by the {\it Chandra} absolute astrometry uncertainty, which is $1.0^{\prime\prime}$ (90\%) for ACIS-I observations according to {\it Chandra} Proposers' Observatory Guide\footnote{\url{https://cxc.harvard.edu/proposer/POG/html/index.html}}. The absolute astrometry of the {\it HST}/ACS images is tied to the {\it Gaia} reference frame, with a typical absolute pointing accuracy of $\sim 10$ mas, negligibly small compared to that of {\it Chandra}.

\begin{figure*}
\centering

\begin{minipage}[t]{0.46\textwidth}
    \centering
    \includegraphics[width=\linewidth]{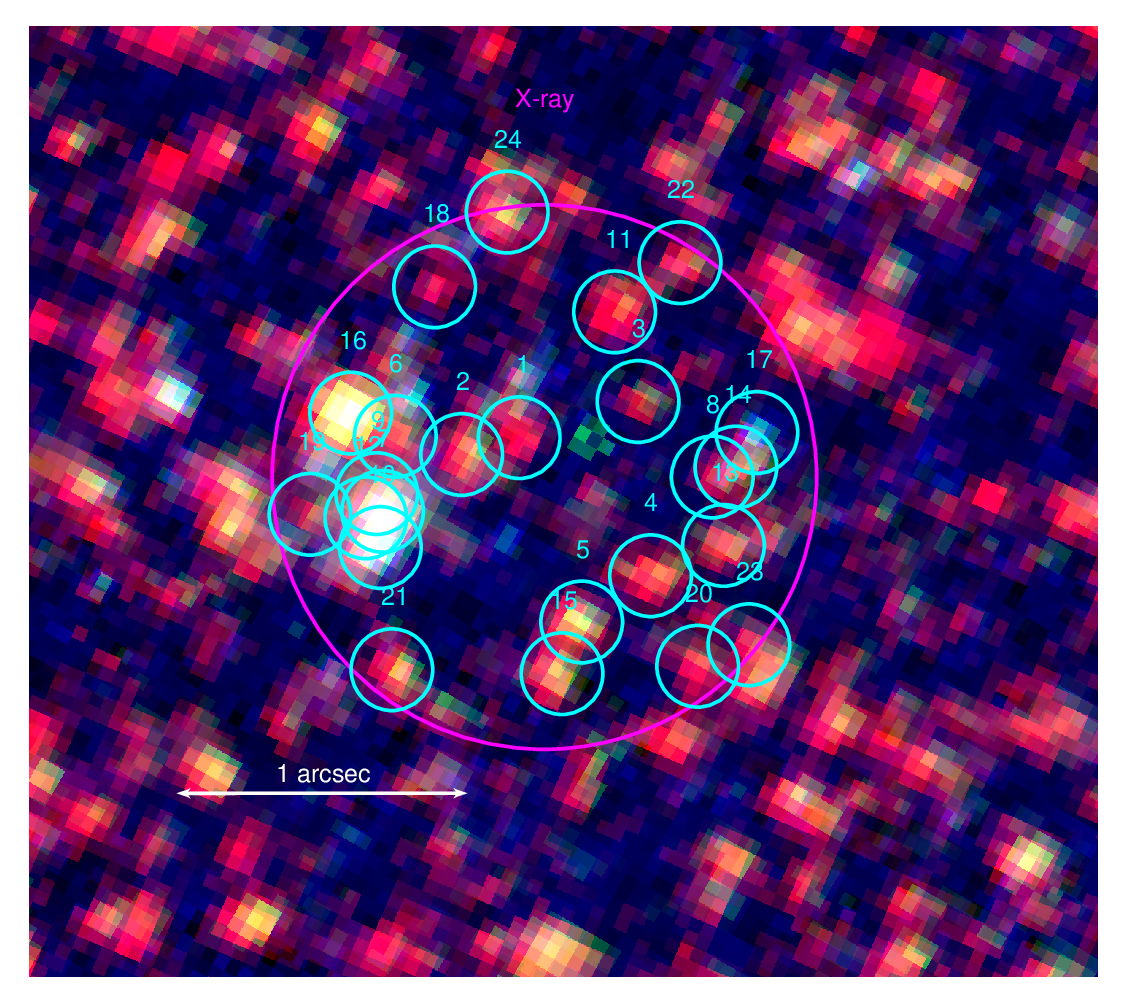}
\end{minipage}%
\begin{minipage}[t]{0.54\textwidth}
    \centering
    \includegraphics[width=\linewidth]{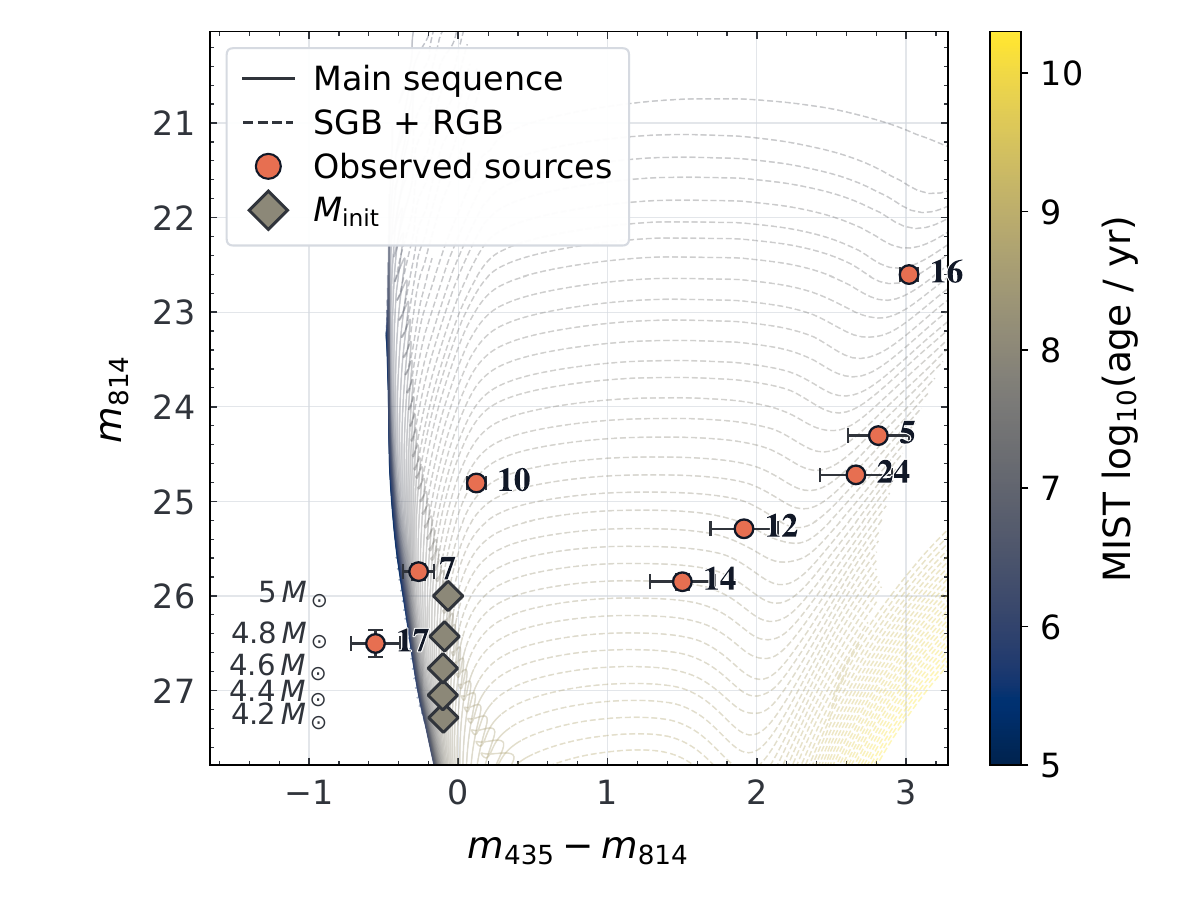}
\end{minipage}
    \caption{
    A multi-wavelength view of the field around M81F4.
    {\it Left panel}: The zoomed-in HST/ACS RGB image of the same field (F814W in red, F606W in green, and F435W in blue).
    In both panels, 
    the large magenta circle marks the X-ray source centroid with a radius of $1^{\prime\prime}$.
    The small yellow circles mark the bright  point-like sources detected in the F435W image, each with a radius of $0.15^{\prime\prime}$. 
    {\it Right panel}: The F435W-F814W versus F814W color-magnitude diagram for these candidates. Only the 8 sources with valid measurements in both bands are plotted, with their ranks labeled. The error bars show the formal $1\sigma$ \texttt{DOLPHOT} photometric uncertainties. The curves show the MIST v1.2 evolutionary models for $0.5-50~M_{\odot}$ at solar metallicity and $v/v_{\rm crit}=0.4$, shifted using a distance modulus of 27.80 and Galactic foreground extinction. Color denotes stellar age, and line style denotes evolutionary phase. The markers and accompanying labels along the $\log_{10}(t/{\rm yr})=8.0$ isochrone indicate the corresponding initial stellar masses, $M_{\rm init}$.}
    \label{fig:hst}
\end{figure*}

We select all optical source candidates within $1^{\prime\prime}$ of the X-ray centroid of M81F4 that have a good-star detection with an apparent magnitude of $m \leq 26$ in at least one of the F435W, F606W, and F814W bands. 
Combining these selections and associating detections across the three bands using a matching radius of $0.1^{\prime\prime}$ yields 24 independent optical source candidates (Figure~\ref{fig:hst}). 
The source positions and apparent magnitudes are listed in Table~\ref{tab:hst}, in order of increasing angular separation from the X-ray centroid, ranging from $0.17^{\prime\prime}$ to $0.98^{\prime\prime}$. Using a cross-band matching radius of $0.1^{\prime\prime}$, 8, 21, and 24 candidates have retained measurements in F435W, F606W, and F814W, respectively. Their apparent magnitudes span $24.93-27.38$ mag in F435W, $23.65-27.92$ mag in F606W, and $22.60-26.50$ mag in F814W. We also searched for but found no {\it Gaia} DR2 \citep{2016A&A...595A...1G} counterpart for any of these sources, probably due to their faintness.

Visual examination of the {\it HST} images reveals no sign of dust extinction features in the vicinity of M81F4, which is consistent with the lack of intrinsic absorption suggested by the X-ray spectra.
Assuming only Galactic foreground extinction ($A_{\rm F435W}=0.324$, $A_{\rm F606W}=0.217$, and $A_{\rm F814W}=0.142$; \citealp{2009ApJS..183...67D}), even the brightest candidate (source \#10 in Table~\ref{tab:hst}), with $m_{\rm F435W}=24.93$ mag, has an absolute magnitude of only $M_{\rm F435W}\simeq-3.19$ mag at the distance of M81.

In principle, one may try to infer the spectral type or stellar mass from single-star isochrone fitting. However, in an accreting X-ray binary, the optical emission may contain contributions from both the donor star and the accretion flow, in particular X-ray reprocessing in the outer disk, such that the observed broadband colors need not follow those of an isolated star. This effect can be especially important for luminous X-ray binaries and may bias any stellar classification based on only three optical bands \citep{2011ApJ...737...81T}. 
We therefore only report the observed photometric properties of the 24 candidates without assigning unique stellar parameters. Nevertheless, in the right panel of Figure~\ref{fig:hst}, we present a F435W-F814W color-magnitude diagram, in which the 8 sources detected in both bands are compared to the MIST v1.2 stellar evolutionary models \citep{2016ApJS..222....8D}.

None of these sources can presently be identified as a secure optical counterpart of M81F4 on the basis of the available HST imaging alone. 
Nevertheless, based on the measured magnitudes, we can safely conclude that no luminous optical source such as a supergiant or a main-sequence star with a spectral type earlier than B0V ($\sim15\rm~M_\odot$) is present in the vicinity of M81F4. 
On the other hand, the photometry of several sources (such as \#5 and \#12) are consistent with a moderately bright subgiant or red giant whose initial mass is about $4-5\rm~M_\odot$. We consider any of these sources a plausible counterpart of M81F4, serving as the donor to an accreting compact object (Section~\ref{subsec:NS}).

\begin{table*}
\centering
\begin{tabular}{ccccccc}
\hline
Rank & Separation & RA & Dec & F435W & F606W & F814W \\
 & (arcsec) & (deg) & (deg) & (mag) & (mag) & (mag) \\
\hline
1 & 0.17 & 148.850989 & 69.166137 & \nodata & $27.92 \pm 0.27$ & $25.05 \pm 0.05$ \\
2 & 0.32 & 148.851154 & 69.166120 & \nodata & $26.34 \pm 0.07$ & $25.16 \pm 0.05$ \\
3 & 0.44 & 148.850648 & 69.166174 & \nodata & $26.74 \pm 0.11$ & $25.76 \pm 0.08$ \\
4 & 0.53 & 148.850612 & 69.165997 & \nodata & $26.76 \pm 0.10$ & $24.42 \pm 0.03$ \\
5 & 0.55 & 148.850810 & 69.165949 & $27.12 \pm 0.20$ & $25.50 \pm 0.04$ & $24.31 \pm 0.03$ \\
6 & 0.57 & 148.851344 & 69.166139 & \nodata & $26.90 \pm 0.12$ & $25.08 \pm 0.05$ \\
7 & 0.61 & 148.851381 & 69.166062 & $25.48 \pm 0.06$ & $25.09 \pm 0.07$ & $25.74 \pm 0.08$ \\
8 & 0.62 & 148.850435 & 69.166097 & \nodata & \nodata & $25.80 \pm 0.09$ \\
9 & 0.62 & 148.851398 & 69.166080 & \nodata & \nodata & $25.68 \pm 0.08$ \\
10 & 0.66 & 148.851387 & 69.166025 & $24.93 \pm 0.05$ & $24.85 \pm 0.03$ & $24.81 \pm 0.04$ \\
11 & 0.66 & 148.850715 & 69.166266 & \nodata & $27.36 \pm 0.21$ & $24.22 \pm 0.03$ \\
12 & 0.67 & 148.851429 & 69.166056 & $27.20 \pm 0.22$ & $27.14 \pm 0.17$ & $25.29 \pm 0.06$ \\
13 & 0.70 & 148.850403 & 69.166027 & \nodata & $26.57 \pm 0.09$ & $25.08 \pm 0.06$ \\
14 & 0.70 & 148.850367 & 69.166107 & $27.35 \pm 0.20$ & $26.86 \pm 0.11$ & $25.85 \pm 0.09$ \\
15 & 0.72 & 148.850866 & 69.165897 & \nodata & $25.99 \pm 0.07$ & $24.80 \pm 0.06$ \\
16 & 0.75 & 148.851472 & 69.166163 & $25.62 \pm 0.06$ & $23.65 \pm 0.01$ & $22.60 \pm 0.01$ \\
17 & 0.80 & 148.850307 & 69.166143 & $25.95 \pm 0.08$ & $26.31 \pm 0.07$ & $26.50 \pm 0.14$ \\
18 & 0.81 & 148.851231 & 69.166291 & \nodata & \nodata & $25.55 \pm 0.06$ \\
19 & 0.87 & 148.851589 & 69.166059 & \nodata & $26.87 \pm 0.12$ & $25.96 \pm 0.09$ \\
20 & 0.89 & 148.850477 & 69.165904 & \nodata & $26.81 \pm 0.11$ & $25.24 \pm 0.06$ \\
21 & 0.90 & 148.851354 & 69.165901 & \nodata & $26.17 \pm 0.06$ & $24.78 \pm 0.04$ \\
22 & 0.93 & 148.850528 & 69.166316 & \nodata & $26.69 \pm 0.14$ & $25.27 \pm 0.05$ \\
23 & 0.97 & 148.850330 & 69.165926 & \nodata & $26.92 \pm 0.12$ & $25.64 \pm 0.08$ \\
24 & 0.98 & 148.851023 & 69.166368 & $27.38 \pm 0.24$ & $25.74 \pm 0.05$ & $24.72 \pm 0.03$ \\
\hline
\end{tabular}
\caption{
The 24 optical source candidates selected by requiring a photometrically qualified detection with $m \leq 26$ in at least one of the {\it HST}/ACS F435W, F606W, and F814W bands within $1^{\prime\prime}$ of the X-ray centroid of M81F4. Sources are listed in order of increasing angular separation from the centroid.
The photometry was obtained from the {\it HST}/ACS images using \texttt{DOLPHOT}. 
Only measurements passing the adopted photometric quality cuts are reported. 
A value of `\nodata' indicates that the source has no retained detection in the corresponding band.
The quoted uncertainties are the formal $1\sigma$ \texttt{DOLPHOT} photometric errors. }
\label{tab:hst}
\end{table*}

%
\facilities{HST(ACS), Swift(XRT), CXO, XMM-Newton, NuSTAR}

\software{Astropy \citep{2013A&A...558A..33A,2018AJ....156..123A,2022ApJ...935..167A}, 
CIAO \citep{2026arXiv260514144F},
DOLPHOT \citep{2016ascl.soft08013D},  
HEASoft \citep{2014ascl.soft08004N}, 
SAS \citep{2004ASPC..314..759G}, 
XSPEC \citep{1996ASPC..101...17A}, 
MIST \citep{2026ApJS..283...41B}}


\end{document}